\documentclass[letterpaper,twocolumn,10pt]{article}
\usepackage[dvipsnames]{xcolor}
\usepackage[T1]{fontenc}
    \usepackage{textcomp}
    \usepackage[utf8x]{inputenc}
\usepackage{fontawesome5}

\usepackage{usenix-2020-09}

\usepackage{tikz}
\usepackage{amsmath}
\usepackage{adjustbox}
\usepackage{lipsum}
\RequirePackage{array}
\RequirePackage{longtable}
\RequirePackage{hyperref}
\RequirePackage{multirow}
\RequirePackage{wrapfig}

\newlength{\fixedCol}
\newcolumntype{L}[1]{>{\raggedright\let\newline\\\arraybackslash\hspace{0pt}}p{#1}}
\newcolumntype{M}[1]{>{\raggedright\let\newline\\\arraybackslash\hspace{0pt}}m{#1}}
\newcolumntype{C}[1]{>{\centering\let\newline\\\arraybackslash\hspace{0pt}}p{#1}}
\newcolumntype{R}[1]{>{\raggedleft\let\newline\\\arraybackslash\hspace{0pt}}p{#1}}

\newcommand{\boldcaption}[2]{{\caption{\textbf{\large #2}\label{#1}}}}
\newcommand{\mystrut}{\rule{0pt}{2.5Ex}}
\newcommand{\TableHead}[1]{\mystrut\textbf{#1}}

\RequirePackage{ifthen}

\newboolean{ShowComments}
\setboolean{ShowComments}{true}  % change to true to show remaining colorful author comments. 
\ifthenelse{\boolean{ShowComments}}%
	{
		\newcommand{\ColorComment}[3]{%
				{\colorbox{#1}{\color{White}   \textsf{\textbf{#2}}} \textcolor{#1}{#3}}}%  Colorful box, initials, phrase 

	}%
	{
		\newcommand{\ColorComment}[3]{}%  Do nothing at all

	}%

\definecolor{rdvcolor}{rgb}{0,0.5,0}\newcommand{\rdv}[1]{\ColorComment{rdvcolor}{rdv}{#1}}
\definecolor{satohcolor}{rgb}{0.5,0,0.5}
\definecolor{michalcolor}{RGB}{255,127,80}
\definecolor{shotacolor}{rgb}{0,0,1}
\definecolor{cocoricolor}{RGB}{238, 130, 238}\newcommand{\cocori}[1]{\ColorComment{cocoricolor}{cocori}{#1}}
\definecolor{naphanncolor}{RGB}{112, 51, 173}\newcommand{\naphann}[1]{\ColorComment{naphanncolor}{whit3z}{#1}}
\definecolor{touchcolor}{RGB}{45,0,134}
\definecolor{shigeyacolor}{RGB}{198,53,39}

\DeclareUnicodeCharacter{6771}{\tou}

\newcommand{\nodelist}[2]{\begin{wrapfigure}[6]{r}{0pt}
\includegraphics[width=0.1\textwidth]{#2}
\end{wrapfigure}
\noindent\hangindent=0.7cm{\bf #1}
}
\begin{document}
%-------------------------------------------------------------------------------

%don't want date printed
\date{}

\title{\Large \bf A Quantum Internet Architecture}

%\rdv{Author list (to be removed for double blind review): rdv, cocori, whit3z, kaaki, michal, satoh, shota, shigeya, Joe Touch}

%for single author (just remove % characters)
\author{
{\rm Rodney Van Meter}\\
Keio University
\and
{\rm Ryosuke Satoh}\\
Keio University
\and
{\rm Naphan Benchasattabuse}\\
Keio University
\and
{\rm Takaaki Matsuo}\\
WIDE Project
\and
{\rm Michal Hajdu\v{s}ek}\\
Keio University
\and
{\rm Takahiko Satoh}\\
Keio University
\and
{\rm Shota Nagayama}\\
Mercari, Inc.
\and
{\rm Shigeya Suzuki}\\
Keio University
%\and
%{\rm Joe Touch}\\
%Keio University
% copy the following lines to add more authors
% \and
% {\rm Name}\\
%Name Institution
} % end author
% \date{\today}
\maketitle

%-------------------------------------------------------------------------------
\begin{abstract}
%-------------------------------------------------------------------------------
%Not more than 200 words, if possible, and preferably closer to 150.
Entangled quantum communication is advancing rapidly, with laboratory and metropolitan testbeds under development, but to date there is no unifying Quantum Internet architecture. We propose a Quantum Internet architecture centered around the Quantum Recursive Network Architecture (QRNA), using RuleSet-based connections established using a two-pass connection setup. Scalability and internetworking (for both technological and administrative boundaries) are achieved using recursion in naming and connection control.  In the near term, this architecture will support end-to-end, two-party entanglement on minimal hardware, and it will extend smoothly to multi-party entanglement and the use of quantum error correction on advanced hardware in the future. For a network internal gateway protocol, we recommend (but do not require) qDijkstra with seconds per Bell pair as link cost for routing; the external gateway protocol is designed to build recursively.
The strength of our architecture is shown by assessing extensibility and demonstrating how robust protocol operation can be confirmed using the RuleSet paradigm.
%Simulations have shown that statistical multiplexing can work well for networks of a small number of endpoints, which we are working to confirm at global internetwork scale. 
%Overall, the combined structure of routing, multiplexing and resource management remains as the most urgent open question. 
%\rdv{Get something implemented and tested...} 
% touch: addressed scale ambiguity by talking about number of endpoints
\end{abstract}

%-------------------------------------------------------------------------------
\section{Introduction}
%-------------------------------------------------------------------------------

% ^^^^6771 ^^^^4eac
% should be Tokyo
% \char"6771
%\touc
%\touc
% \symbol{"6771}
%\unichar{"00E4}
%\unichar{"20AC}
%\unichar{"6771}\unichar{"4EAC}
% should be the atom symbol
%\unichar{"269B}
   
%\rdv{Check figures at \url{https://github.com/sfc-aqua/quisp/blob/f9db91abded6ff78e5e9160c342a0c41fd728404/doc/Rule\%20Engine.md}
%}

The coming Quantum Internet will provide new encryption services, enhance the sensitivity of sensor networks, and couple distant quantum computers to enhance secure computation, share quantum data and increase the size of problems that can be attacked~\cite{Wehner18:eaam9288,van-meter14:_quantum_networking,I-D.irtf-qirg-principles,kimble08:_quant_internet}. Hardware components are in rapid development~\cite{PRXQuantum.2.017002}. Numerous architecture and protocol factors have also been investigated, but not yet brought together into a coherent architecture~\cite{aparicio11:repeater-muxing,van-meter:qDijkstra,PhysRevA.93.042338,dahlberg2019link,I-D.van-meter-qirg-quantum-connection-setup,kozlowski00:qnp}.  And yet, our decades of experience with the classical Internet clearly show that architecture and hardware must develop in tandem, and that of the two architecture matures more slowly.  Thus, it is imperative to begin laying the foundation for an architecture, driving development of hardware and learning from proposed applications as we go.

It is important to recognize that there will be an internetwork, a network of networks~\cite{van-meter12:_q_internetworking}.
Without a doubt there will be more than one network architecture;
but to build a true Quantum Internet there will ultimately have to be only a single internetwork architecture.

\subsection{Quantum Communication is Different}

%\rdv{Both the nature of the information, and the challenges posed in addressing them, especially errors.}

We can summarize quantum communication as follows: \emph{nonlocality is the goal, teleportation is the heart, decoherence is the reality, and the speed of light is still the constraint.}

\emph{Quantum entanglement} arises from quantum nonlocality,
a phenomenon in which distant systems obeying quantum mechanics share a state, allowing them to demonstrate correlations as if they are in direct, seemingly instantaneous communication.  Entangled states can be either bipartite or multipartite.

Teleportation is currently the heart of quantum networking~\cite{bennett:teleportation}, as it is the primary method of transferring quantum information encoded in physical quantum states.  In quantum teleportation, the state of a quantum variable is destroyed in one place and reconstructed in another.  Teleportation from network node $A$ to node $B$ consumes a special entangled state spanning $A$ and $B$, known as a \emph{Bell pair}; hence, the task of a quantum network is to continually produce enough end-to-end entanglement to satisfy applications.  Moreover, a form of teleportation known as \emph{entanglement swapping} is used to stretch link-level entanglement into end-to-end entanglement. Other types of quantum networking, e.g., involving superposition but not teleportation, appear to be limited to single-hop configurations and are thus not considered further here.

Unfortunately, quantum data is exceedingly fragile.  Photons get lost, so generally speaking we must use acknowledged link layers (though there are exceptions), dramatically affecting throughput. Errors in quantum states caused by noise, imperfect control of memories, etc. are collectively called \emph{decoherence}. One measure of decoherence suffered is \emph{fidelity},
an estimate of the closeness between the actually-achieved and desired quantum states.
% touch: talked about classical later two paragraphs later, as distinct from this "acknowledged link layer"
% an estimate of the probability that our state is still good.

Finally, although entanglement shows nonlocal correlations, it cannot be used to communicate faster than the speed of light.
Essentially, all quantum communications require supporting classical communication, which is naturally limited to $c$. Measurement outcomes on entangled qubits are (anti)correlated and at a first glance may appear to violate special relativity. However, the measurement collapse is random and cannot be controlled, making faster-than-light communication impossible.
% touch: added michael's text here as the last sentence to address my concern
% \rdv{Figure?}

All quantum communication relies on a classical communication infrastructure to enable control and coordination. This classical infrastructure is a distinct communication system that operates at the application layer, similar to how some routing protocols run as an application to manage router forwarding tables. This classical network need not share paths or topology with the quantum network it manages, but necessarily interconnects every controllable quantum network component, whether quantum (e.g., teleportation repeater) or classical (e.g., optical switch).

To read this paper, readers need only the notions above, along with the general idea that we are working with \emph{qubits}, quantum binary digits that can be entangled with each other and follow a few simple rules~\cite{divincenzo2000piq}. Qubits can be encoded into photons (using a variety of encoding methods) or stored in stationary memories (implementable in many different physical systems). For a brief summary of quantum information concepts and both popular and technical references, see Appendix~\ref{sec:qconcepts}.

%\lipsum[7]

\subsection{Quantum Communication is Desirable}

The unusual characteristics just described would be little more than a curiosity (or a physics experiment) without compelling reasons to integrate quantum communications into our existing IT ecosystem to provide new or better services. We can divide applications into three main, overlapping areas: cryptographic services, sensor networks, and distributed quantum computation~\cite{I-D.irtf-qirg-quantum-internet-use-cases,van-meter14:_quantum_networking,cuomo2020towards,q-proto-zoo}.

The best-known quantum cryptographic service is \emph{quantum key distribution} (QKD), in which quantum characteristics are used to assess the probability of the presence of an eavesdropper as a stream of shared, random bits is created~\footnote{Roughly speaking, QKD can be done using single photons~\cite{bennett:bb84,Pirandola:20,RevModPhys.92.025002} or E2E entangled states~\cite{PhysRevLett.68.557,ekert1991qcb}.  Single-photon demonstration networks have existed since the early 2000s~\cite{elliott:qkd-net}, but without the ability to store and manipulate states mid-path, they are single-purpose networks and do not provide E2E security; instead, they depend on classical relays with only hop-by-hop security. Here, we focus on more general, entanglement-based systems.}. These random, shared, believed-to-be-secret~\cite{ekert2014,portmann:2102.00021,RevModPhys.92.025002} bits can be used in key cryptographic protocols~\cite{Alleaume201462,elliott:qkd-net,mink09:_qkd_and_ipsec}. However, this is not the only cryptographic service that is possible; secret sharing~\cite{crepeau:_secur_multi_party_qc,hillery1999qss,PhysRevA.59.162,markham2008graph}, secure election protocols~\cite{Tani:2012:EQA:2141938.2141939}, and byzantine agreement protocols~\cite{taherkhani18:qba,ben-or2005fast} are all known.

The second category, sensors, encompasses a range of uses. Arguably, QKD itself is a sensor application, as it physically detects the presence or absence of an eavesdropper. Other uses include enhanced interferometry for telescopes~\cite{PhysRevA.100.022316,PhysRevLett.109.070503} and higher-precision clock synchronization~\cite{komar14:_clock_qnet,ilo-okeke2018remote}, both of which can be viewed as using entanglement as a form of reference frame for time and space~\cite{PhysRevLett.91.217905,1367-2630-16-6-063040,RevModPhys.79.555,PhysRevLett.74.1259,PhysRevLett.87.167901}. Challenges include determining whether the required precision for classical control of the quantum elements exceeds the gains from the use of entanglement in practice, and the extremely high data rates (entanglement generation rates) required.

%\rdv{Not sure this is the right set of references; maybe Buhrmann? My IEEE Computer instead of my book?}
The final area is distributed quantum computation~\cite{I-D.irtf-qirg-quantum-internet-use-cases,cuomo2020towards,buhrman03:_dist_qc,raz1999esq}, where individual quantum processors are networked together, communicating and sharing their resources to carry out quantum information processing tasks in a coordinated way.
Extension of the paradigm of delegated quantum computation leads to applications such as blind quantum computation~\cite{broadbent2009universal,fitzsimons2017private}, where a client is able to delegate her computation to a quantum server without revealing information about its input, the computation itself or its output.

\subsection{Quantum Repeaters}
\label{sec:repeaters}
%\rdv{The four roles: base entanglement, entanglement extension, error management, participating in the network}

\begin{figure}[!t]
    \centering
    % \adjustbox{margin=1em,width=\linewidth,set height=4cm,set depth=4cm,frame,center}{Quantum repeaters}
    \includegraphics[width=\columnwidth]{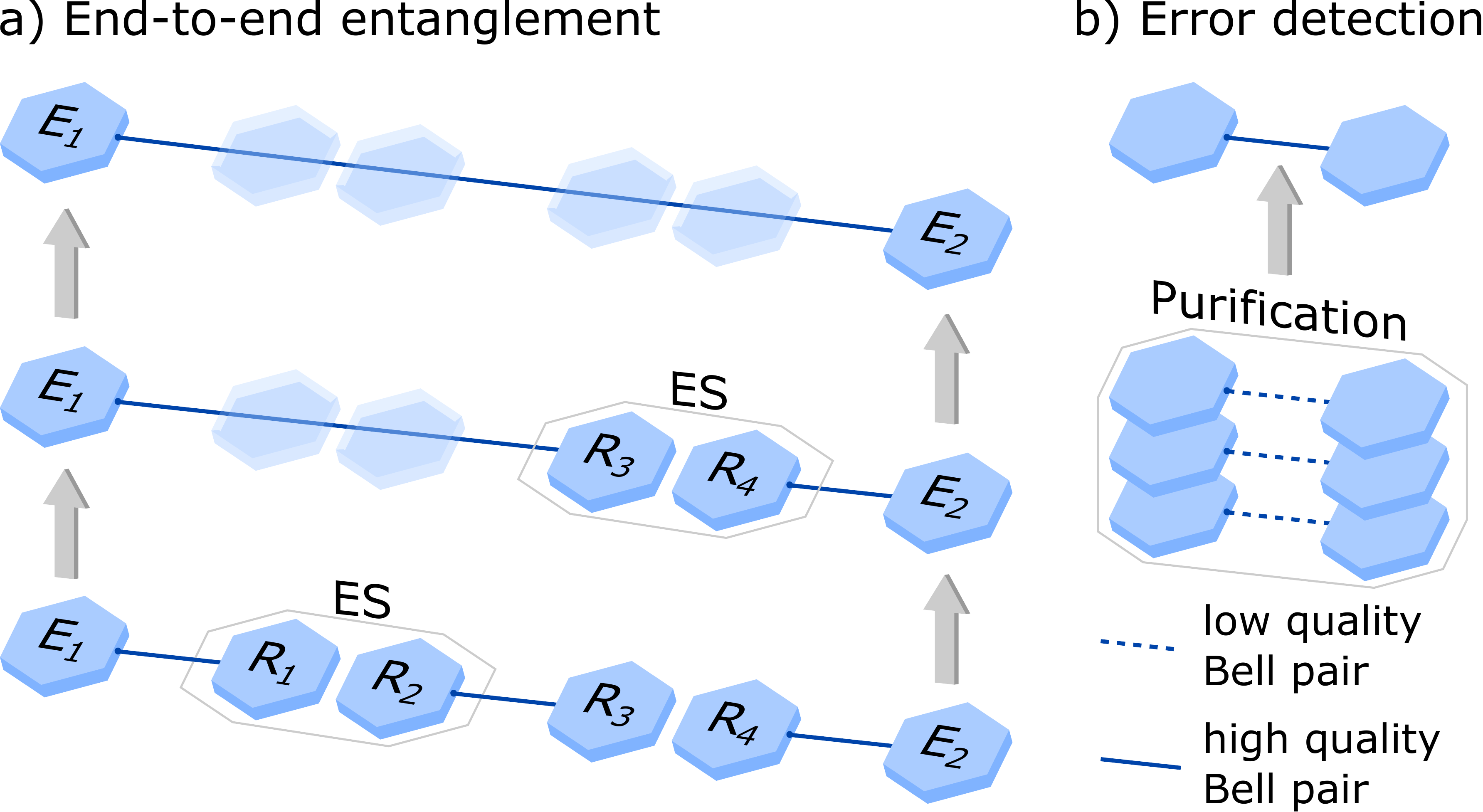}
    \caption{Quantum repeaters build end-to-end distributed entanglement for use by applications at end nodes.  In the basic form shown in (a), that process is a distributed computation, depending on \emph{entanglement swapping} (ES) to lengthen entanglement to span multiple hops and a form of error detection, shown in (b), known as \emph{purification}, where multiple low quality Bell pairs can be winnowed down to a single pair of higher quality through a testing protocol that consumes some pairs.
    %\shota{the definition of link seems different from others} \rdv{I agree w/ Shota. What term should we use for entangled states or edges in an entangled graph? I don't like link, I prefer to keep that for a physical link.} \michal{I think for now we can just stick to Bell pairs so the last could read ``..., where multiple low quality Bell pairs can be winnowed down to a single Bell pair of higher quality''. And we can think about the appropriate terminology when it comes to graph states later since we do not really talk about them here.}
    }
    \label{fig:fig1-repeaters}
\end{figure}

Quantum repeaters are very different from classical signal repeaters; quantum states cannot be amplified or simply regenerated~\footnote{Quantum amplifiers~\cite{caves2012quantum,chia2019phase} are an existing quantum technology capable of boosting certain quantum signals, however quantum states where this is possible have limited use in the context of quantum communication~\cite{chia2020phase}.}, and as a general rule cannot be faithfully copied. Instead, the work of the network is to perform a distributed computation that builds the end-to-end entanglement that applications consume. Repeaters and routers serve as waypoints in that E2E problem, and perform four main tasks:

\begin{enumerate}
    \item {\bf Creating base entanglement:} Typically using single photons (though there are exceptions to this rule~\cite{devitt2016high}), neighboring repeaters entangle stationary memory qubits. The most common outcome of this process is a \emph{Bell pair}. 
    A number of different link architectures can be used to achieve this task~\cite{jones16:comm-pro}.
    \item {\bf Entanglement extension:} Achieved via \emph{entanglement swapping}~\cite{PhysRevLett.71.4287} shown in Fig.~\ref{fig:fig1-repeaters}(a), two entangled Bell pairs, $A\leftrightarrow B$ and $B\leftrightarrow C$ can be spliced to form a single $A\leftrightarrow C$ Bell pair. Classical communication is required.
    \item {\bf Error management:} Loss of photons is handled using acknowledged link layers, but state errors and operation (gate) errors must be addressed as well; \emph{purification} is a form of error detection, shown in Fig.~\ref{fig:fig1-repeaters}(b). With enough resources and high enough basic fidelity, quantum error correction can be used.
    \item {\bf Network operations:} Nodes must monitor their own links as well as participate in routing, multiplexing, network operational security, etc. in both networks and internetworks.
    Our use of this term includes what might be considered both the control and management planes of the quantum network, both of which operate over a classical network that interconnects quantum devices at the classical application layer.  This is the focus of this paper.
\end{enumerate}

The most commonly discussed architecture uses purification and entanglement swapping; unless otherwise stated, in this paper we are discussing these first generation, or 1G, networks. Purification requires bilateral confirmation of a qubit measurement result; on even parity, the entangled state is kept and proceeds, while on odd parity the state must be discarded. Entanglement swapping transfers entanglement from one node to another, which requires communicating with two nodes, one of which may be required to adjust its state using information known as a \emph{Pauli frame correction}.  Coordination of these operations in a robust but maximally asynchronous fashion is one of the primary tasks of the network protocol.

\begin{figure}[!t]
    \centering
\includegraphics[width=\linewidth]{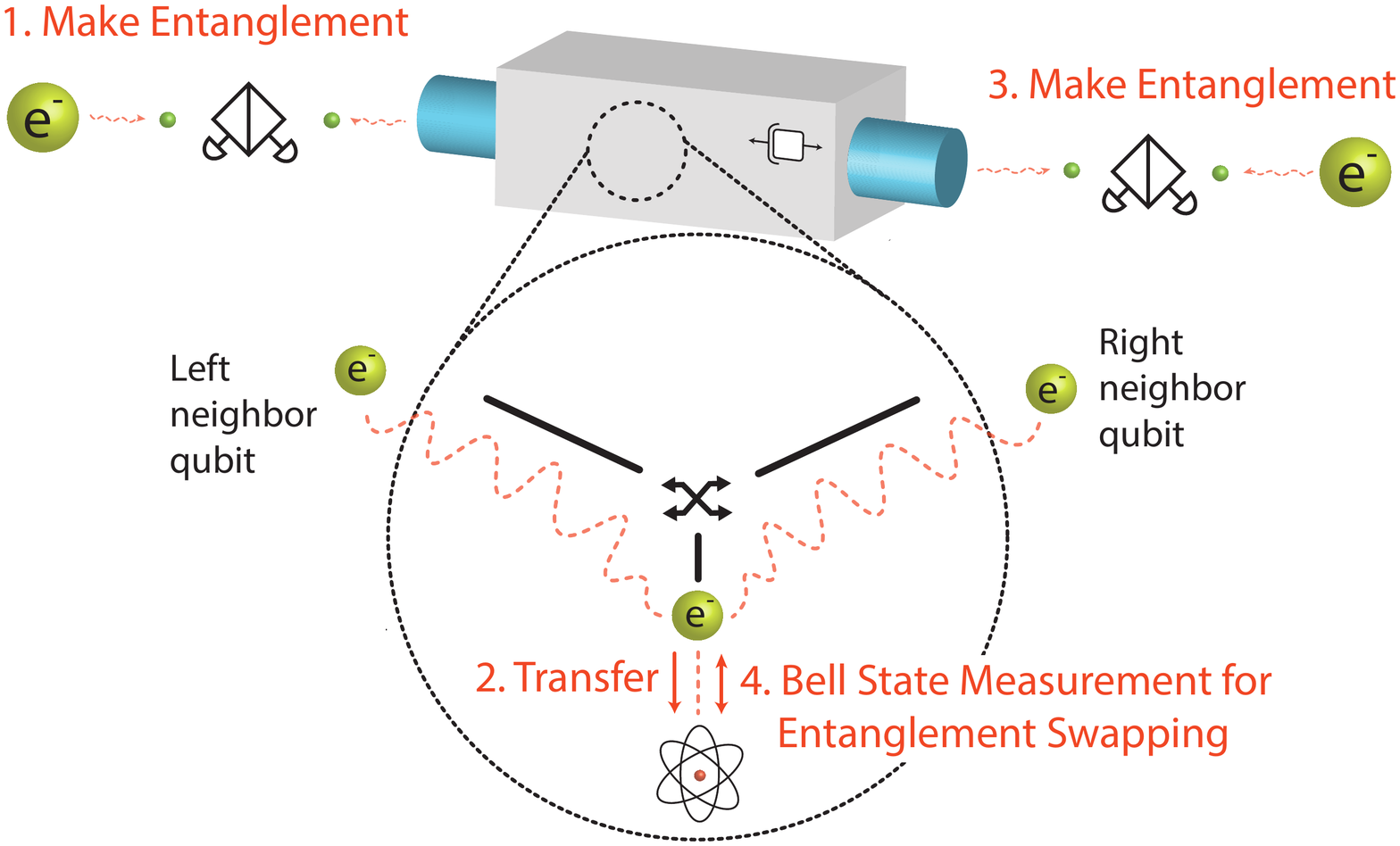}
     \caption{Present-day quantum repeaters~\cite{pompili2021realization} represent the absolute minimal form of hardware: a single transceiver qubit ($e^-$), a single buffer memory qubit (atom symbol \faAtom), a two-port optical switch in front, and the ability to initialize, store, manipulate and measure the qubits. This repeater can only attempt to build entanglement to either the left or the right in a given cycle; e.g., after succeeding in making entanglement to the left (Step 1), then the transceiver qubit's state is transferred to to the buffer qubit (Step 2), and entanglement to the right is attempted (Step 3).  Once entanglement to the right is achieved, entanglement swapping is performed via a Bell state measurement (joint measurement) of the two qubits (Step 4).  This is followed by classical communication with the neighbors (Step 5, not shown).}
    \label{fig:2port-minimal}
\end{figure}
\begin{figure}[!t]
    \centering
    \includegraphics[width=\linewidth]{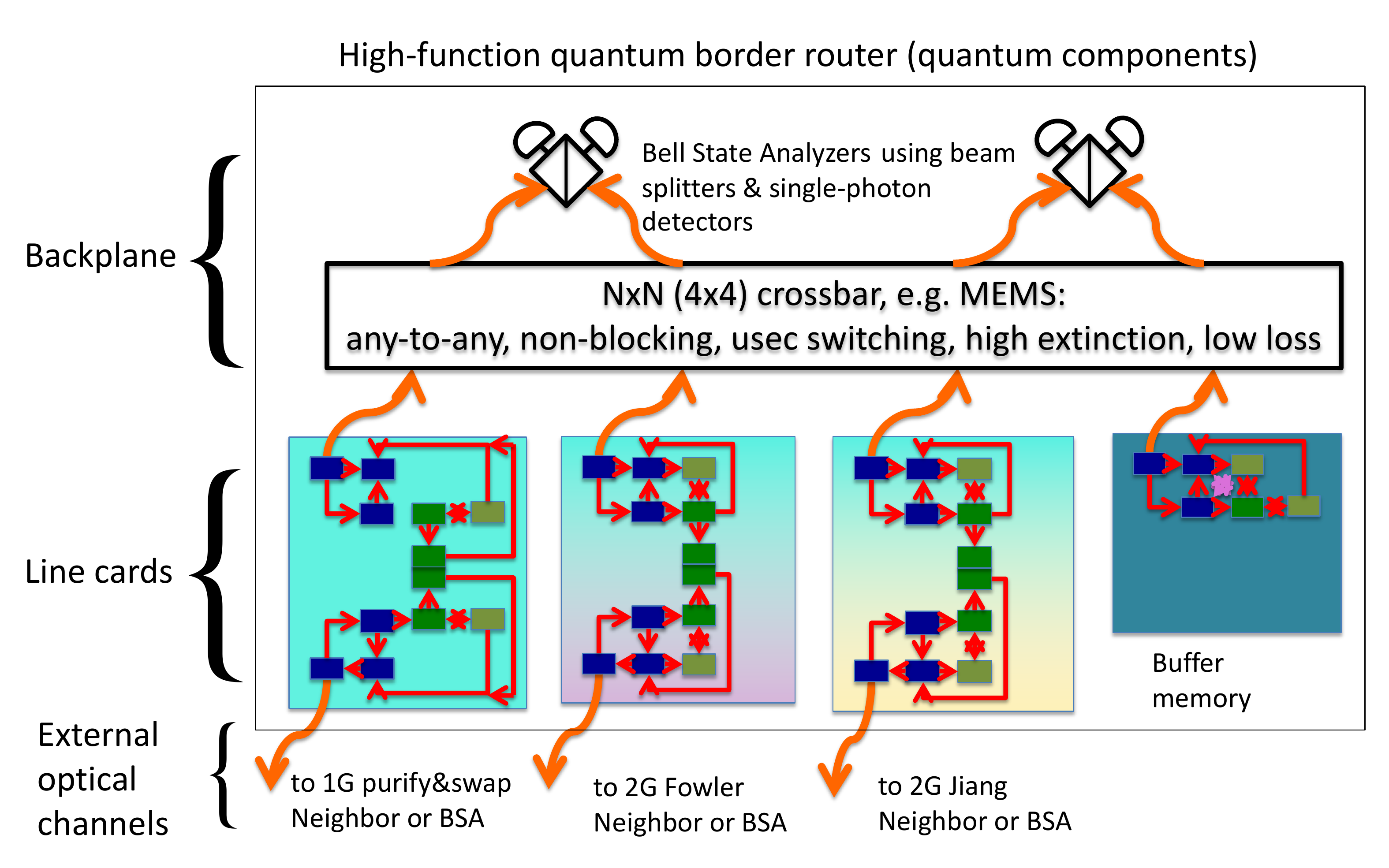}
    \caption{A full quantum router with hardware architecture similar to today's commercial Internet routers will have QNICs (line cards) coupled via a backplane consisting of optical ports, an optical switch, and Bell State Analyzer measurement devices.  Using the BSAs, the qubits in the backplane buffers at the top of the line cards are entangled while the transceiver qubits in the lower portion attempt to create entanglement with neighboring nodes.  Once both backplane and neighbor entangled states are made, entanglement swapping is used within each line card to splice the long-distance entanglement.
    %The marked flow within each line card represents changing software roles for qubits as they attempt entanglement, then become entangled, then are used to facilitate end-to-end entanglement. 
    A number of steps in hardware complexity (and cost) will exist between the minimal configuration of Fig.~\ref{fig:2port-minimal} and this one. %\rdv{Text needs to be bigger.}
    %\touch{any way to put most of the detail in the body and have the figure caption just be a caption? this applies throughout} \rdv{This one might be a little long, but my preferred caption style has evolved over the years. I prefer a descriptive caption that tells us what we are looking at, rather than just "Quantum Router Architecture". Science and Nature go with much longer captions. What is the key thing you want the reader to learn from the figure? This also helps reviewers, who are likely flipping through the paper rather than reading it end to end.}
    }
    \label{fig:full-router}
\end{figure}

%\subsection{Goals of an Architecture}

\subsection{Architecture Decision Points}
\label{sec:decisions}

In developing a Quantum Internet architecture, our goals are similar to those of the classical Internet: we want a system that is robust in operation; easy to implement; and meets requirements such as scalability, security, manageability, and autonomy.
Good definitions of interfaces will allow subsystems and hardware implementations to evolve independently and systems will continue to interoperate over time spans of (human) generations.
Because we are designing an internetwork, our goal is to create a homogeneous service over heterogeneous subpaths, however, this must be balanced against the fact that early hardware generations will have substantial differences in capabilities.
%\touch{I think quantum *Internet* should also have a goal of creating a homogeneous service over heterogeneous subpaths, just like the Internet; otherwise, you have network but not an Internetwork} \rdv{Mmm, my immediate reaction was, "Joe's right," but then I thought a little more. One of the challenges in QI evolution is likely to be that available services are initially heterogeneous.}

A number of key design decisions must be made:
%\michal{I have slightly changed the wording and formatting of all the points below, please check.}

\begin{enumerate}
\item \emph{The nature of the fundamental service.}
Is it Bell pairs, measured-out classical bits, qubit teleportation or multipartite graph states? (Sec.~\ref{sec:service})
\item \emph{The nature of connections.} Is the network 1G, utilizing entanglement swapping and purification?
Or is it 2G/3G~\cite{muralidharan2016optimal}, establishing connections using quantum error correction (QEC)?
Alternatively, the connections can be all-photonic, without quantum memories \cite{azuma2015all,hilaire2021resource}. (Sec.~\ref{sec:semantics})
% entanglement swapping and purification (1G), or QEC (2G, 3G)? (affects internetworking)
\item \emph{APIs.} How do applications access the services provided by the network? What is a socket for quantum communication? (Sec.~\ref{sec:sockets})
\item \emph{Conveying requests.} The protocols for achieving the above services must be designed, including naming conventions for quantum resources. (Sec.~\ref{sec:rulesets}.)
\item \emph{Stateful connections.} Connections will require both quantum and classical state at each repeater along a path, at least as long as that component is actively participating in building quantum states for the endpoints.
%For example, after teleportation has completed, the neither quantum nor classical state at a given repeater may no longer be needed.  
What sort of handshake/signalling mechanism is used to establish a connection? Is this centralized or distributed? (Secs.~\ref{sec:two-pass}, \ref{sec:muxing} and \ref{sec:recursion})
%\item \emph{Establishment of connections.} What sort of handshake/signalling mechanism is used to establish a connection? Is this centralized or distributed? (Sec.~\ref{sec:two-pass})
\item \emph{Node types.} The state of technology determines the types of nodes we can build; the above items determine the types of nodes required to build a quantum network. (Sec.~\ref{sec:node-types})
\item \emph{Routing.} How do we pick a path or route through the network? (Sec.~\ref{sec:routing})
\item \emph{Multiplexing discipline for resources}. 
Options for multiplexing the use of quantum resources may include circuit switching, time division muxing, statistical muxing or buffer space muxing. % Regardless of method, the focus here is on ways the quantum system itself is multiplex; there can be additional classical multiplexing, e.g., of the optical path between nodes as well.
Naturally, stateful connections and many of the muxing candidates require authentication, authorization and accounting. (Secs.~\ref{sec:muxing},~\ref{sec:aaa})
\item \emph{Security.} Quantum networks allow numerous new attack vectors which have to be considered~\cite{satoh2021attacking}. These attacks sometimes coincide with the defining property of the service provided by the quantum network, e.g., as in QKD; in other cases, such as for distributed computation, they represent challenges to be overcome. (Sec.~\ref{sec:security}) 
\item \emph{Making an internetwork.} 
How should the networks come together to create an internetwork and what is the nature of their interactions? (Sec.~\ref{sec:recursion})
\end{enumerate}

The above list is by no means exhaustive but covers the critical points.
For a more complete list, the interested reader can turn to the QIRG Internet Draft~\cite{I-D.irtf-qirg-principles}.

After proposing answers to these questions in the next several sections, we provide some evidence for the correctness of our choices (Sec.~\ref{sec:evidence}) before concluding (Sec.~\ref{sec:conclusion}).

% \rdv{Paper outline referred to above; complete it here. Don't forget to describe what the appendices hold.}

\section{Quantum Network Services}
\label{sec:service}

\subsection{Semantics: Bell Pairs and On-Path\\ Distributed Computation}
\label{sec:semantics}

%\rdv{Pause to think: what should a quantum network \emph{do}?}

%\rdv{Don't forget, time is part of the service.}

Entanglement is the resource that will fuel quantum applications such as QKD, teleportation, quantum sensing, or delegated quantum computation.
Continuous, reliable and efficient replenishment of this resource is one of the primary tasks of a quantum network.
However, entangled states come in many shapes and sizes~\cite{horodecki2009quantum,hajdusek2010entanglement}.

Bell pairs are the most basic bipartite states and form the fundamental building blocks of entangled quantum networks.
They can be generated by a network link equipped with stationary quantum memories at each end that are entangled via flying photons.
Due to the Bell pairs' importance to virtually all quantum communication protocols it is generally agreed that they will be part of the fundamental network service.

% Individual links use memories at each end and photons to entangle. It seems pretty obvious to me that the fundamental primitive is the physical Bell pair generated by the link (which we also call a "base entangled state").  Everything else builds on top of this.

End-to-end multipartite states such as GHZ, W and graph states~\cite{hein2004multiparty} are resources for a variety of multiparty protocols, and therefore are likely to be extremely valuable.
Bell pairs alone would be sufficient; multipartite states can be built using them, but because the efficiency will matter, it is an open question whether multipartite states are part of the fundamental service or should be created and managed entirely by applications running at end nodes.
% (In particular, I'm not at all sure what the APIs at the responders are like to make something like this happen.  What is {\tt listen()} like in this case?)

%\michal{}
For this reason, we focus on distribution of Bell pairs in this manuscript.
This distribution can be achieved in a number of ways and depending on the nature of the connection networks are classified into three generations~\cite{muralidharan2016optimal}.
1G quantum networks build E2E entangled pairs using physical Bell pairs, spliced and error tested using entanglement swapping and purification. Such connections are the most basic way of establishing E2E entanglement and therefore the first implementations of quantum networks are expected to be 1G. 2G and 3G networks are designed to reach and maintain higher fidelities (especially useful for distributed computation) by utilizing quantum error correction, placing very demanding requirements on the hardware. These error management schemes require work at each repeater, but with the goal of E2E error management~\cite{nagayama21:e2eep}.

%\michal{}
All of the above generations of quantum networks rely on quantum memories to store the qubits while the networks entangle them using flying photons.
It is prudent to mention that all-photonic quantum repeaters have been proposed that do not require quantum memories~\cite{azuma2015all,hilaire2021resource}.

In Sec.~\ref{sec:repeaters}, we noted that this distribution of Bell pairs is, in fact, a limited form of distributed computation all along the path. The semantics of the network service must be defined to take this into account.

Finally, it is important to note that \emph{time is part of the service}~\cite{I-D.irtf-qirg-principles}. For sensor applications in particular, very high precision timestamps of some events are necessary information, and must be provided to applications.

% However, that's made more complicated by the possibility of all-optical repeaters~\cite{azuma2015all,hilaire2021resource}.  An especially tricky issue is how to terminate such connections and how to make all-optical connections interoperate with other types of connections/nodes/networks.

% Same for multipath connections.  It's a pretty obvious idea, and sorry but I can't remember who first proposed them in print (Perdrix? Benjamin? Bruss?).  My own opinion is that the benefits of multipath are likely to be minor, as a. often, the first hop will be the bottleneck anyway, b. asymmetry in the paths in the real world means that benefits will be minimal, and c. I assume there will be a lot of competition for resources on the net, and so the occasions when you can actually acquire the resources to do multipath effectively will be few.  Oh, and d. the software complexity is high.  So, I think it's doable in QRNA+RuleSet, but it's far down my list of things to work on. 
% \touch{Multipath is one of many optimizations that can be contained within the service provided by a virtual link. This is discussed further in Section [REF?] ??}

% So, let's stick with Bell pairs for the moment as both the fundamental link service and the *primary* end-to-end service.
% QRNA and RuleSet-based approach can handle multiparty states as we develop it over time.

\subsection{Application Access: Quantum Sockets}
%\subsection{Application Access to Network Services: Quantum Sockets}
\label{sec:sockets}
Once entangled qubits are ready, an application consumes them. As mentioned earlier, there are three types of currently envisioned applications. It is possible to categorize the three into two types: those that use qubits in larger quantum applications and those that measure the qubits to produce classical information right away. Applications that consume qubits directly will measure the result immediately after the execution of the application; thus, eventually, both cases have a classical result.

We are designing the Quantum Socket API (API from now on) in an analogy with the classical socket API widely used in the classical Internet. The API is very similar to the classical socket API. Like the classical socket API, the API has functionalities such as: creating a socket, connecting to the socket endpoint, reading, writing, setting options, and destroying the socket. The API is node-type agnostic; i.e., it can handle three end-node types (MEAS, COMP, SNSR described in Sec. \ref{sec:node-types}) corresponding to three different application classes. 

Operations on nodes of the type that return classical information (MEAS, SNSR) as described earlier are synchronous since the result will eventually become classical reads (read system calls, etc.). For these applications, the stochastic arrival time of completed Bell pairs is not a problem. %In contrast, direct manipulation of qubits (COMP) is asynchronous in its reception. 
In contrast, COMP nodes involve substantial coordination with other work at a quantum computer. How to build distributed quantum programs that deal robustly with stochastic entanglement delivery, e.g. via asynchronous callbacks, is an open problem.
%Therefore, we need to execute the application which consumes the qubit on the reception of the useable qubit asynchronously.
%Thus, for COMP nodes, we use RuleSets to execute the application code via QCIRC action. Since QCIRC action includes MEAS action at the end of execution, the result will be classical.

Both applications or controlling programs can configure specific parameters for each physical interface, such as the RuleSet specific to the QNIC, via the ioctl-like interface. In other words, classical components communicate with physical quantum components via the socket API.

%\rdv{How do apps interact w/ these services?  Cite valley fold, B, C, T.}

% \michal{We don't really mention Application Access as a separate Point in Section 1.5. Should we?} \rdv{done.}

\section{Expressing Connection Semantics: RuleSets}
\label{sec:rulesets}

% We have addressed how routers at the boundary can make entangled states that cross the borders. 
%\touch{As with the Internet, a quantum switch is assumed to interconnect like quantum services and technologies, whereas a quantum router is an interoperability-enabling device capable of interconnecting different quantum services and technologies.??}

Having just established that the core service of a quantum internet is building E2E entanglement, now we need an internetwork protocol capable of communicating the actions necessary to span different connection architectures (1G, 2G). Here, we describe a mechanism efficient enough for use as the basis of a network protocol, and rich and abstract enough for use as an internetwork protocol (Sec.~\ref{sec:recursion}).

1G networks need a mechanism for conveying requests such as, ``Bob, once you get a Bell pair with Alice and a Bell pair with Charlie, execute entanglement swapping, then send the Pauli frame correction to Charlie and a notice-of-entanglement-transfer to Alice,'' and ``If you have two Bell pairs with Alice, both with fidelity less than 0.9, then purify.'' 2G networks will work on \emph{logical} qubits encoded using quantum error correction, making for complex operations for entanglement swapping and error correction while presenting high-fidelity logical qubits to applications.

Our approach is to define Rules that have a \emph{condition clause} and an \emph{action clause}, very analogous to the OpenFlow extensions of classical software defined networking (SDN)~\cite{mckeown2008openflow}.  For a connection, each node is given a RuleSet that should comprehensively define what to do as local events occur (entanglement success, timeout, etc.) and as messages arrive~\cite{matsuo2019quantum,matsuo2019simulation}.
This RuleSet-based operation is the heart of our work, and allows for explicit reasoning about how to achieve the maximum asynchrony and autonomy in the network (rather than waiting for explicit instructions at every operation or attempting to make everything proceed in lockstep). Our version of RuleSets includes local state, which cannot be expressed in SDN OpenFlow.

%\michal{I rewrote the following paragraph. It was referring to QRNA which we have not defined yet. This is because initially, we discussed recursion before the RuleSets.}
%Furthermore, the RuleSet approach meshes well with recursion-based internetworking, discussed in Sec.~\ref{sec:recursion}.
%Here, RuleSets distributed to virtual nodes of a network can be easily expanded at the border routers provided they are equipped with rewrite engines.
%Here, we refer to the structural recursion of RNA applied as a forwarding mechanism. In structural recursion, an encapsulated network is represented as a virtual (possibly multipoint) link or virtual router (which are the same in RNA), in which the endpoints of the encapsulated network are coincident with the endpoints of the virtual link/router.

There is one RuleSet for each connection. Once a resource (e.g., link level Bell pair) is assigned to a RuleSet, that assignment does not change. How that assignment is done is the responsibility of the multiplexing scheme (Sec.~\ref{sec:muxing}).

RuleSets and any qubits at the nodes that are currently assigned to a particular connection are  \emph{connection state} that must be held at each repeater/router. The scalability of this needs to be assessed, and it affects AAA (Sec.~\ref{sec:security}), but we currently see no approach to quantum networking that allows mid-path routers and repeaters to be fully stateless.

% \michal{These \emph{stages} are different from network levels used when discussing recursion. This can lead to confusion. Which one should we rename?}
% \naphann{If using the stage-based RuleSet would explain RuleSet as}

%\rdv{The problem with the term "Level" is that it has been used before to talk about how many hops the entanglement spans.}

%\naphann{I also think that maybe the name RuleSet might not blend well with the Level-block, but we definitely need to come up with a better name than Level. I'll make a figure to illustrate it for the RuleSet pic, after everyone gets a clear view of it maybe we can get a better name}

We have adapted the RuleSet approach from~\cite{matsuo2019quantum,matsuo2019simulation} and incorporated an additional construct called a \emph{Stage}. A RuleSet can be thought of as a program that oversees the processing of the states.  Entangled states are allocated to a given Stage. In each Stage, there can be multiple Rules.
Each Stage can also have its own variables which are shared by Rules in the same Stage.
After one of the Rules in the Stage fires, the entangled state is either promoted to the next Stage or declared defunct and the physical resources are returned to the pool available for reuse.
%, which can be thought of as threads running inside the same process which share the same resource pool.
%The RuleSet can have variables which are shared by all Rules.
%, the Rules can read from and write into these shared variables. \michal{The following was part of the sentence but seems like a comment: (this should be handled by some semaphores? same concept as concurrent variables)} 
%Qubits are allocated to each Stage. Stages have their own ordering, where lower Stages are allowed to move their qubits to higher Stages but not the other way around.
This ensures that the flow of qubits is unidirectional and terminates with being either delivered to an application or service, or discarded (either consumed as part of the protocol operation or determined to likely be in error).

%\naphann{We can remove the RuleSet variables which are shared by all Rules but this would lead to greedy allocation of resources and RuleSet with lower priority might not get allocation of resources until higher priority RuleSet finished. This wouldn't be a problem for the whole circuit switch multiplexing? As of right now I don't have other solutions to avoid this problem.}
%\touch{not sure what you mean by "Level" here; is this recursive structural embedding? or protocol layering for multilink integration?} 
%\naphann{The stage is a block where multiple Rules reside in it. So RuleSet contains stage-Block, and the stage-Block contains Rules. Rule is the smallest element. stage-Block is created to support branching. They are not the same Level in the RNA sense. Maybe this should be rename to something else}

\subsection{Naming States (Qubits)}
\label{sec:naming}
%\cocori{I'm not quite sure what should I write here, but I guess I should write the way of identification and control of the qubit.}
Managing the qubits and agreeing on the consumption of entangled states among shared nodes are two of the most critical tasks for RuleSets.
In the IP architecture, network addresses are associated with a network interface; here, we assume the same. Thus, a physical qubit can be uniquely identified by its network address and the index of the qubit within the QNIC, using the tuple \verb|<QNICAddress,QubitIndex>|. QNIC firmware applies quantum operations based on that index, and the tuple is unique within the scope of the network address.

However, rather than this physical address, we are usually interested in the state (e.g., half of a Bell pair) that is held in the qubit, which is dynamic and has a finite lifetime; the distinction is philosophically similar to a register versus a temporary variable.  Therefore, when nodes that share entangled resources want to communicate changes to other parties, they use another (external) name which is only known by the shared parties. This name needs to be unique. Initially, the name is determined by one of the nodes involved in the creation of the link-level Bell pair (e.g., the BSA node described in the next section). The name might be, for example, the tuple \verb|<NodeAddress,Timestamp>|, where the timestamp is of high enough precision that at most one Bell pair may have been created. The mapping of that external name to internal qubit address is maintained independently and privately by each node.

When entanglement swapping is completed, a new name for the Bell pair is created by the node performing the swapping. That name is communicated to the two end points as part of the notification of the entanglement being transferred to new partners.

\subsection{Messages}

% \rdv{Enumerate the messages that can be received across the network. (n.b.: this is not the Omnet++ messages, which include local IPC.)}

Tab.~\ref{tab:messages} lists the primary messages included in our protocol. (Naturally, each message transmission is initiated by an Action clause, and its reception matches a Condition clause; in the interest of space these are not included in Tabs.~\ref{tab:conditions} and ~\ref{tab:actions}.) Purification involves testing the parity of two qubits at each end and exchanging the results using a measurement outcome (MEAS) message. Each end compares the parity it calculates to the parity it receives, and either discards both Bell pairs (on mismatch) or raises the software's estimated fidelity of one and discards the other (on match).  Entanglement swapping requires that both ends be notified of the transfer of entanglement to new partners, and one end must also receive a Pauli frame update.

\newcommand{\FIRSTcategory}[1]{\multicolumn{6}{|l}{\TableHead{#1}}}
\newcommand{\category}[1]{\\\hline\FIRSTcategory{#1}}
\newcommand{\component}[1]{\\\cline{2-6}&\TableHead{#1}}

\begin{table*}[!t]
% \begin{longtable*}
\boldcaption{tab:messages}{Protocol Messages}
\setlength{\fixedCol}{1.2in}
%\begin{adjustbox}{width=\textwidth, height=\textheight/2}
%\begin{tabular}{p{0.1in}L{1.0in}|L{1.4in}|L{0.55in}c|L{3.6in}|}
\begin{tabular}{p{0.1in}L{0.75in}|L{1.2in}|L{0.75in}c|L{2.8in}|}
%\begin{tabular}{p{0.1in}L{1.3in}|p{\fixedCol}|L{\fixedCol}|L{0in}L{3.5in}|}
%\commonheadings
	\hline\hline
	\multicolumn{2}{|l|}{\TableHead{Name}} &
	\TableHead{Descriptive Name} & 
	\TableHead{Arguments} & 
	%\TableHead{\scriptsize Milestone}   % delete for now
	&
	\TableHead{Comments} \\\hline\hline
\FIRSTcategory{Remote Events (Message Transmission)}
\component{FREE}
& Release a state
& Partner addr., resource IDs
& 
& Release a state back to the free pool. Used after purification.
\component{UPDATE}
& State change notification
& Partner addr., resource IDs, Pauli frame correction
& 
& Used to indicate a Pauli frame correction to a state. Most commonly used with TRANSFER to complete entanglement swapping.
\component{MEAS}
& Measurement outcome
& Partner addr., resource IDs, result
& 
& Exchange purification results. Each partner sends this message, and a separate rule will recognize whether purification results agree and proceed appropriately. Numerous types are possible.
\component{TRANSFER}
& Entanglement transfer notification
& Partner addr., resource IDs %, Pauli frame correction
& 
& Distribute the result of a swapping circuit. Generalizes to a notice of entanglement transfer from one location or partner to another. Carries a new resource ID to used for the resulting state.
%\component{SWAP}
%& Entanglement Swapping % (also known as Entanglement Transfer)
%& Partner address, resource identifier(s), Pauli frame correction
%& 
%& Distribute the result of a swapping circuit. This generalizes to a notice of entanglement transfer from one location or partner to another.
%\cocori{SWAP could be a different name} %\naphann{Can we just say send correction message or notification message? If we use SWAP then we can't use this action for teleport or GHZ correction which I think is almost the same thing and we probably don't want to keep adding new actions for each message type?}
%
%\component{PUR}
%& Purify Entanglement
%& Partner addr., pur. type, pur. outcome
%& 
%& Purify entanglement; numerous types are possible. Each partner sends this message, and a separate rule will recognize whether purification results agree and proceed appropriately. %\cocori{Should we have multiple purification types as different actions?}
%
\\\hline
\end{tabular}
%\end{longtable*}
%\end{adjustbox}
\end{table*}

\if0
\naphann{As of right now I think the general classes of messages are as follows:}
\begin{enumerate}
    \item correction/notified messages; after some other nodes finished performing its end of the contract like swapping / teleport. Which the receiving nodes must perform some corrections or update its resource table or maybe discard the qubits. \naphann{this should capture all the swapping / teleport / GHZ correction}
    \cocori{Why the BSA result is not included here?}
    \item purification message; should include the qubits and type \cocori{Also partner address? Is this purification results?}
    \item Bell state measurement timing notification; originated from HoM to two neighbours of the link \naphann{should this be included?} \cocori{Isn't this included in the first one?} \naphann{the first one I envisioned it to be from end node to end node only. This one is originated from HoM}
    \item BSA result
\end{enumerate}
\fi

\subsection{Condition Clauses}

%\naphann{The current kaaki's implementation of RuleSet has variable inside Conditional Clauses which means that checking it multiple times can lead to some unexpected side effect. I don't think this is a good idea. The Condition Clause checking procedure shouldn't have side effects and if we want to apply side effects it should be done via the Action clause.}

%\naphann{I think instead of saying "trigger for moving from one state to another". It should only be referred to as arrow, since it can also repeat the same state like when measurement occurred but still not enough for the tomography RuleSet. Also since every Rule is independent and can run in parallel, maybe it's not the same as state machine. I misunderstood RuleSet at the start. We probably better describe it as the condition to execute a programs while the actions is the actual program to run.} \rdv{That's pretty close to what I had in mind: Condition to run a small program, and the program itself. But with very limited semantics and runtimes.}

%\newcommand{\FIRSTcategory}[1]{\multicolumn{6}{|l}{\TableHead{#1}}}
%\newcommand{\category}[1]{\\\hline\FIRSTcategory{#1}}
%\newcommand{\component}[1]{\\\cline{2-6}&\TableHead{#1}}

\begin{table*}[!t]
% \begin{longtable*}
\boldcaption{tab:conditions}{Condition Clauses}
\setlength{\fixedCol}{1.2in}
\begin{tabular}{p{0.1in}L{0.6in}|L{1.2in}|L{0.75in}c|L{3.1in}|}
%\begin{tabular}{p{0.1in}L{1.3in}|p{\fixedCol}|L{\fixedCol}|L{0in}L{3.5in}|}
%\commonheadings
	\hline\hline
	\multicolumn{2}{|l|}{\TableHead{Name}} &
	\TableHead{Descriptive Name} & 
	\TableHead{Arguments} & 
	%\TableHead{\scriptsize Milestone}   % delete for now
	&
	\TableHead{Comments} \\\hline\hline
\FIRSTcategory{Local Software Events}

%\component{NOP}
%& No operation
%& 
%& 
%& Used for protocol testing or generating \emph{sui generis} events; matches unconditionally.

%\naphann{same argument as in Action clauses; although for Condition if we specify action that has side effects; it would mean we just testing whether our action works or not}

\component{CMP}
& Check whether a variable is equal, less than, or greater than some values
& variable ID, comparison operator, value
& 
& Used to track number of operations done (e.g. purification count, measurement count, or number of notification message received)
% \naphann{I think we should generalize the PURCOUNT and MEASCOUNT into this one.}
%
\component{TIMER}
& Timer expiration
& Timer ID
& 
& Must be used with caution when dealing with distributed states; race conditions can occur.
%\component{Connection \newline architecture} 
%
%\component{FID}
%& Fidelity
%& Required fidelity %$0.5 < F < 1.0$
%& 
%& The primary ``meets application requirements'' clause for delivering to apps at EndNodes. %\rdv{Could just be a form of RES?} \naphann{I agree that this should be included in the RES if we want the resource to have higher fielity than some value; Or if we specify the argument to and check for \emph{lower than certain value} and stick it in every Level then we can discard qubits if the resource has been sitting in a certain Level for too long, although if we have a way to track it, could we just use TIMER instead?}
%
%\component{MEASCOUNT}
%& Measurement Count
%& 
%& 
%& Used in tomography to track and ultimately terminate operation \rdv{Is this a software hack, or good architecture?}
%
%\component{PURCOUNT}
%& Purification Count
%& 
%& 
%& Used to track number of purification operations \rdv{Is this a software hack, or good architecture?}
%
%\component{Logical interoperability} 
%& RTR
%& Medium
%& 
%& Long-term issue in practice; knowing its feasibility today is reassuring
% \category{Local Hardware Notifications}
% \category{Remote Events (Message Reception)}
\category{Quantum State Events (Local Hardware Notifications, Message Reception)}
\component{RES} 
& Enough Resources 
& Partner address (or wildcard) and fidelity %\naphann{types?}
& 
& Matches Bell pairs. Used commonly for purification and entanglement swapping. Used to check fidelity of Bell pairs, this also serves as the primary ``meets application requirements'' clause for delivering to apps at EndNodes.
%\rdv{this is actually a bad name}
%\naphann{should we include the type of qubits as well? Physical, types of logical qubits?}
%\component{RESLEFT} 
%& Enough Resources Left
%& Medium
%& 
%& Used for two-party entanglement swapping \rdv{Right now, I'm scratching my head as to why this is a separate clause type rather than an AND of a left RES clause and a right RES clause? I think maybe it was some software hack of kaaki's because he had some trouble somewhere?}
%\component{RESRIGHT} 
%& Enough Resources Right
%& 
%& 
%& Used for two-party entanglement swapping
\\\hline
\end{tabular}
%\end{longtable*}
\end{table*}

Tab.~\ref{tab:conditions} shows the Condition Clauses that can be defined in Rules. The Condition Clauses can be thought of as defining the trigger for moving from one state to another in a state machine, while the Action Clause for the Rule defines the side effects.

Sometimes, a Condition Clause needs to match only one entangled state, for example when matching a Bell pair and deciding to deliver it to an application (in which case it passes out of our ken). More often, it needs to match two: either with the same end points, for purification, or with different end points, for entanglement swapping.

\subsection{Action Clauses}

\begin{table*}[!t]
% \begin{longtable*}
\boldcaption{tab:actions}{Action Clauses}
\setlength{\fixedCol}{1.2in}
%\begin{adjustbox}{width=\textwidth, height=\textheight/2}
%\begin{tabular}{p{0.1in}L{1.0in}|L{1.4in}|L{0.55in}c|L{3.6in}|}
\begin{tabular}{p{0.1in}L{0.75in}|L{1.2in}|L{0.75in}c|L{2.8in}|}
%\begin{tabular}{p{0.1in}L{1.3in}|p{\fixedCol}|L{\fixedCol}|L{0in}L{3.5in}|}
%\commonheadings
	\hline\hline
	\multicolumn{2}{|l|}{\TableHead{Name}} &
	\TableHead{Descriptive Name} & 
	\TableHead{Arguments} & 
	%\TableHead{\scriptsize Milestone}   % delete for now
	&
	\TableHead{Comments} \\\hline\hline
\FIRSTcategory{Local Software Actions (Classical)}

%\component{NOP}
%& No operation
%& 
%& 
%& Protocol testing or match without any action.
% \naphann{NOP is used for protocol testing? How do we know if the test work if it doesn't have any side effect / some kind of actions? I think we should remove it? In my proposed Stage-base RuleSet, NOP wouldn't serve any purpose and I think? I also think that if we can implement the "or" conditional either via the Stage-based RuleSet or in some shape or form to kaaki's RuleSet we can remove the variable stored in the Conditional clause altogether and this NOP would not be required}

\component{SETTIMER}
& Set timer
& Timer ID
& 
& Use with caution; distributed race conditions can occur.
\component{PROMOTE}
& Promotion of qubits
& Qubit IDs, Rule ID, Stage
& 
& Used to transfer the control/ownership of Qubits from current Rule (Stage) to Another Rule (Stage)
%
%\component{ENC}
%& Encode physical qubits(resources) into logical ones
%& Qubit identifiers
%& 
%& Used to transform some set of physical qubits into logical one \naphann{In a way, we don't need this we can just transfer / operate on them as a group but it would be easier if we have this(?)}
%
\component{FREE}
& Free qubits 
& Qubit IDs
& 
& Release qubits to the pool of unallocated resources.
\component{SET}
& change value of a Rule/RuleSet variable
& variable identifier
& 
& Can be used to track how many measurements have occurred for tomography.
%
%\component{Logical interoperability} 
%& RTR
%& Medium
%& 
%& Long-term issue in practice; knowing its feasibility today is reassuring
\category{Local Hardware Actions (Quantum)}
\component{MEAS}
& Measure qubits
& Qubit IDs, meas. basis
& 
& Measure one or more qubits in specified basis or a randomly chosen one.
%
%\component{RANDMEAS}
%& Measure qubit in random basis
%& 
%& 
%& Measure in random basis used in the tomography \naphann{How about combining this with MEAS and use "RANDOM" as argument?}
%
\component{QCIRC}
& Apply quantum circuit
& Qubit identifiers, Qcircuit
& 
& Apply a general unitary quantum operation on one or more qubits, without measuring. Bell state measurement, purification, and entanglement swapping execute QCIRC first, then MEAS. Encoding into logical qubits also uses.
\\\hline
\end{tabular}
%\end{longtable*}
%\end{adjustbox}
\end{table*}

Tab.~\ref{tab:actions} shows the Action Clauses that can be defined in Rules.  The Action Clauses can be thought of as defining sequences of local quantum operations and messages to be sent. The actions are chosen from a restricted set of options and do not include loop primitives; despite the existence of QCIRC which applies a quantum circuit, this is not a Turing complete computation platform.  Conditional execution is done by creating separate rules with distinct Condition clauses. These restrictions make it easier to reason about distributed protocol actions in terms of termination, robustness, deadlock, security, and other issues.

As noted above, message generation is not included in this table but is a natural consequence of QCIRC, MEAS and some of the local software actions.

% \input{action_clauses_table}

% \rdv{Not Turing complete, but simple conditionals can be done by taking advantage of the condition clauses.}

\subsection{Two-Pass Connection Setup}
\label{sec:two-pass}

%---------------------------
\begin{figure}[t]
\begin{center}
\includegraphics[width=\columnwidth]{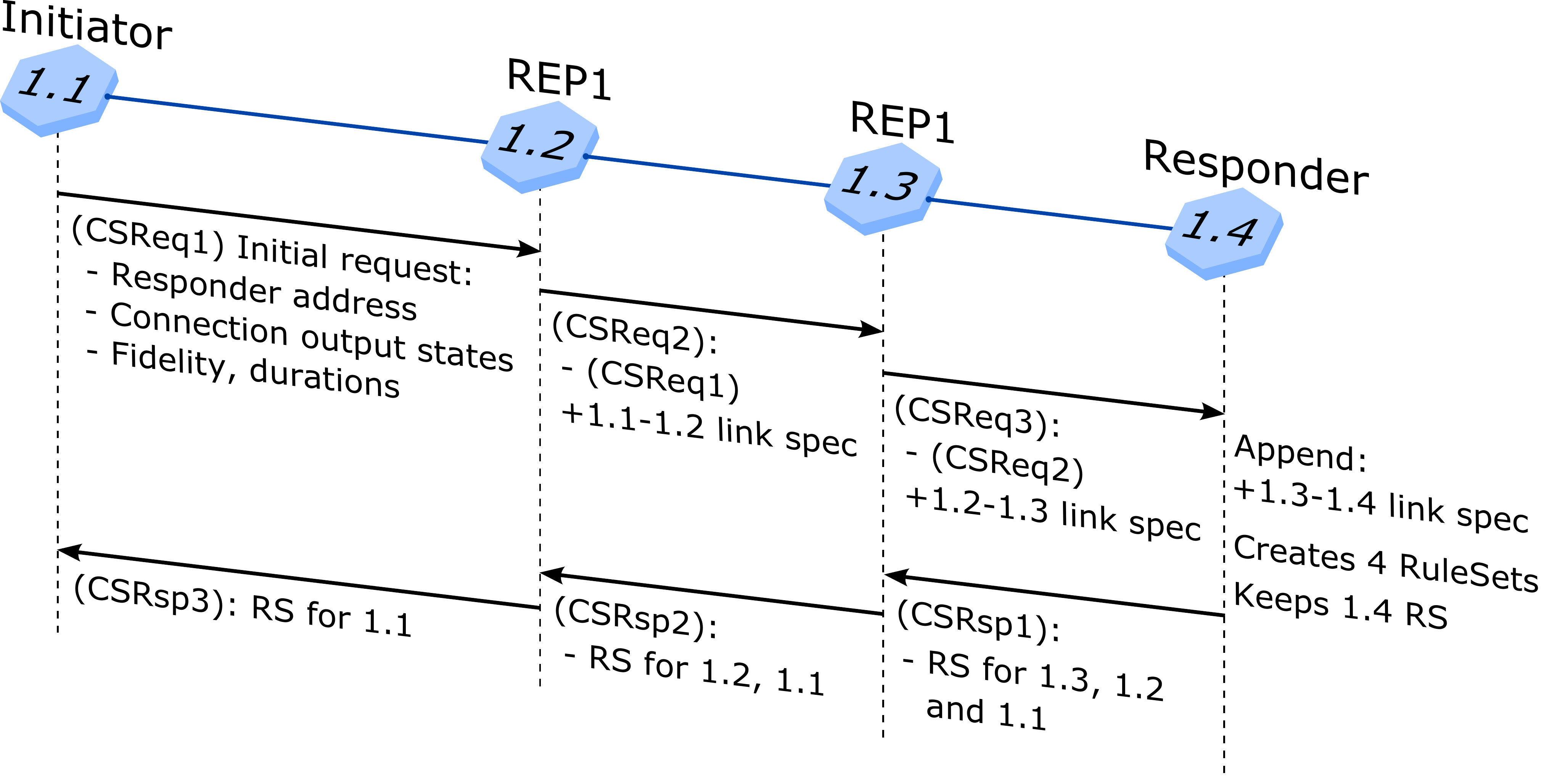}
\end{center}
\caption{\label{fig:simple-setup} Two-pass connection setup (CS) within a single quantum network. RuleSets are created by the Responder, offering a distributed innovation point.}
\end{figure}

Our approach to connection setup uses two passes, as proposed by Van Meter and Matsuo~\cite{I-D.van-meter-qirg-quantum-connection-setup}. On the outbound leg (starting at the \emph{Initiator}), information about links and available resources is collected.  The connection request eventually reaches the \emph{Responder}, which takes that information and builds RuleSets for every node along the path.  Those RuleSets are distributed in a return pass, then the operation for the connection begins.

Setup within a single network is illustrated in Fig.~\ref{fig:simple-setup}. As in the classical Internet, we expect that the majority of connections will be initiated by a client node reaching out to a server. This architecture places the server in charge of RuleSet creation, allowing service providers a single point of innovation; if they create better RuleSets than their competitors, then connections will be faster or more robust, providing a competitive advantage.

%---------------------------
% \begin{figure}
% \begin{center}
% \includegraphics[width=\columnwidth]{simple-setup.pdf}
% \end{center}
% \caption{\label{fig:simple-setup} Two-pass connection setup (CS) within a single quantum network. RuleSets are created by the Responder offering a distributed innovation point.}
% \end{figure}

\section{Networking}

The previous section discussed individual connections in the abstract; here we show how to operate in complex topologies with complex traffic patterns and actors whose interests aren't always perfectly aligned.

\subsection{Quantum Network Components}
\label{sec:node-types}

Quantum networks are distinct from their classical counterparts because they cannot exist in isolation; quantum networks incorporate and rely on classical networks to interconnect their components to enable classical control. So despite the name, a quantum network is really a hybrid of a quantum and a classical network.
%\michal{I don't fully understand the following sentence so I am not sure whether they still fit.} As with a classical network,  a link and a switch are behaviorally equivalent; some switches are internally implemented using link technologies, such as a shared bus or shared memory that emulates a bus.

Just as today's classical Internet consists of Ethernet switches, IP routers of varying capabilities, home routers, WLAN access points, and terminals of various types, nodes comprising the Quantum Internet will come in a variety of flavors.
All of the node types below can be implemented in numerous technologies (NV diamond, ion traps, superconducting, quantum dot)~\cite{ladd10:_quantum_computers}, using a variety of optical qubit representations (polarization, time bin, spatial path, energy/wavelength, etc.).
We divide these into three categories: \emph{end nodes}, \emph{repeater nodes}, and \emph{support nodes}.

End nodes represent hosts that wish to execute a quantum application such as quantum key distribution, secret sharing and blind quantum computation.
The technological maturity required of an end node heavily depends on the desired application.
There are three major kinds of end nodes:
%\begin{description}

%\item[MEAS]
\nodelist{MEAS}{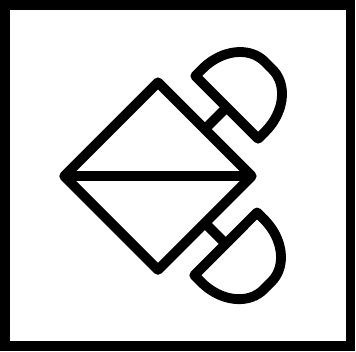}
A node that can only measure received photons (in at least two different ways) and does not store qubits is actually surprisingly useful. A pair of such nodes can conduct quantum key distribution, or a single node of this type can serve as a terminal connecting to a full COMP node in order to execute one form of secure blind quantum computation~\cite{PhysRevA.87.050301}. However, its error management capabilities are very limited.

\nodelist{COMP}{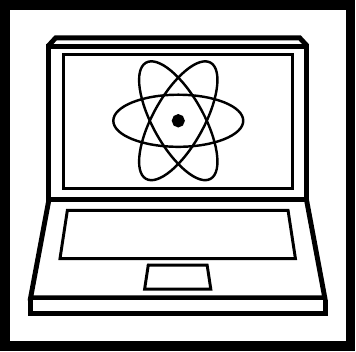}
Computational end node capable of measuring quantum states as well as storing them in a quantum memory. This greatly enhances the nodes functionality and leads to advanced applications such as blind quantum computation~\cite{broadbent2009universal, hayashi2018self}. This node may vary in its processing abilities. Simple clients may be only able to generate, store and manipulate single-qubit states while advanced quantum servers may be able to create large multi-qubit entangled states and hence be capable of universal fault-tolerant quantum computation.

% A full computational end node is a full quantum computer, which is capable of running both standalone quantum programs and fully utilizing the capabilities of a quantum Internet.  Here, we make no distinction as to whether the nodes are NISQ or fully fault tolerant, and whether they are capable of blind computation.  Connecting nodes of the right type is the responsibility of applications. 

% \lipsum[20]
%item[MEM]
% \nodelist{MEM}{icons_MEM.pdf}
% \rdv{I always forget what this particular node is useful for...E91?  But a MEAS can do that, so what else?}
% Memory node. Has the ability to store a quantum state but unlike the COMP node lacks any advanced processing ability.

%\item[SNSR]
\nodelist{SNSR}{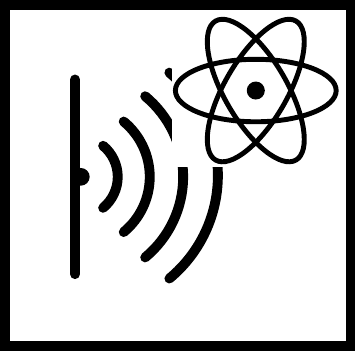}
A sensor node uses the entangled states in a cyber-physical operation, e.g. as a reference frame for interferometry or clock synchronization.  For these nodes in particular, recall that \emph{time is part of the service}.
%\lipsum[20]
%\end{description}

Quantum repeaters are responsible for distribution and management of entanglement across the quantum network.
We have three kinds of repeater nodes:

%\begin{description}
%\item[REP1] 
\nodelist{REP1}{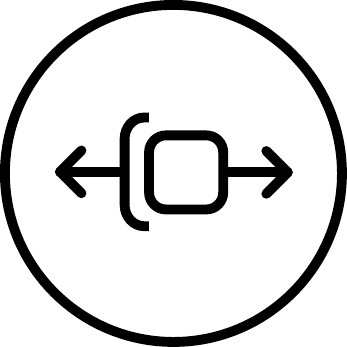}
A 1G repeater.  Always has two interfaces; a recent experiment (Fig.~\ref{fig:2port-minimal} and ~\cite{pompili2021realization}) allows only one to be active at a time, but the generalized form allows both to be active simultaneously.
Its primary task is to perform entanglement swapping and error management in the form of purification on physical qubits.

\nodelist{REP2}{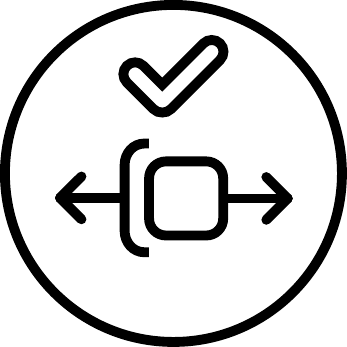}
A 2G repeater. Has the same primary task of entanglement swapping as REP1 but operates at the level of encoded logical qubits composed of multiple physical qubits. Error management is achieved via error correction, signified by the check mark in the REP2 icon. REP2 must be equipped with hardware capable of handling a large number of physical qubits, which necessitates more advanced computational capabilities.

%\item[RTR] 
\nodelist{RTR}{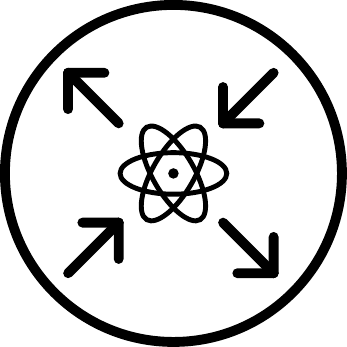}
A router. As in Fig.~\ref{fig:full-router}, a router likely consists of multiple line cards and a backplane, but for network architectural purposes, the important fact is that a router runs a full suite of protocols governing network operations.  Typically, an RTR will have three or more network interfaces, and is capable of governing a network border, where it may be called upon to speak both 1G and 2G protocols and to rewrite RuleSets, behaving as a Responder for connection requests (outbound or transit). 
%\rdv{Arguably, 1G and 2G routers should likewise be recognized as separate node types.}
%\end{description}

Finally, support nodes are tasked with aiding end and repeater nodes in entanglement distribution.
There are five kinds of support nodes:

%\begin{description}

%\item[EPPS] 
\nodelist{EPPS}{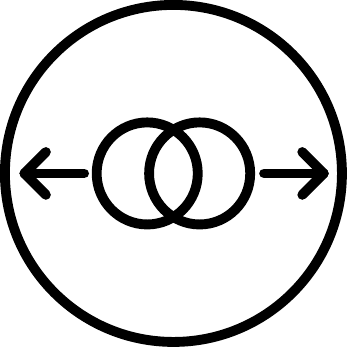}
An entangled photon pair source, implemented using e.g. symmetric parametric down conversion (SPDC).  An EPPS simply produces pairs of entangled photons, which must be captured or measured at link end points.  An EPPS can be used in terrestrial links~\cite{jones16:comm-pro} or on a satellite, with the photons captured by telescopes on the ground~\cite{Yin1140}.

%\item[BSA]
\nodelist{BSA}{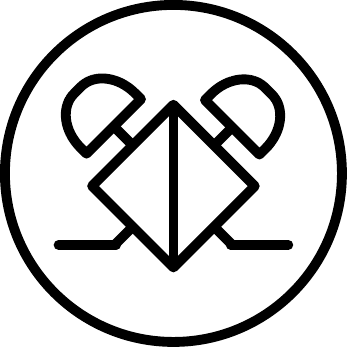}
Bell State Analyzer, which projects two photons into one of the Bell states; usually used to swap memory-photon and photon-memory entanglement to memory-memory entanglement.
The theoretical efficiency limit with linear optics implementation is $50\%$.
The hardware complexity of the BSA depends on the particular qubit encoding.
% \rdv{Accepted edit, but we haven't talked about different qubit representations at all, have we?}
% Some BSAs recognize only a single type of Bell state, others can recognize two. \rdv{But not all four in pure optical, right?}

\nodelist{RGSS}{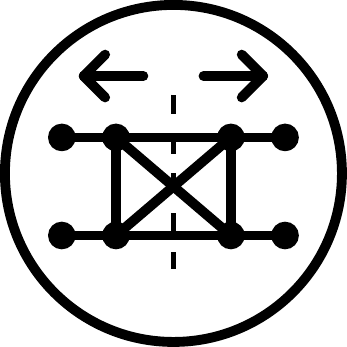}
Repeater Graph State Source generates entangled multipartite photonic states used in memoryless repeater networks.
It sends one half of the generated repeater graph state to its neighboring nodes where the photons are measured.

\nodelist{ABSA}{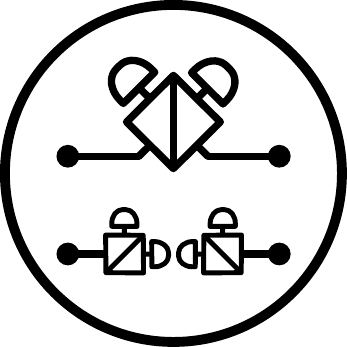}
Advanced Bell State Analyzer.  The basic BSA always performs the same operation, but all-optical repeaters based on repeater graph states require two-photon and single-photon measurements. The measurement basis (type of measurement) is selected dynamically based on prior measurement outcomes as well as the logical encoding and structure of the underlying repeater graph state. 
This makes the hardware, software and protocol implementations much more complex than a BSA.

%\item[OSW]
\nodelist{OSW}{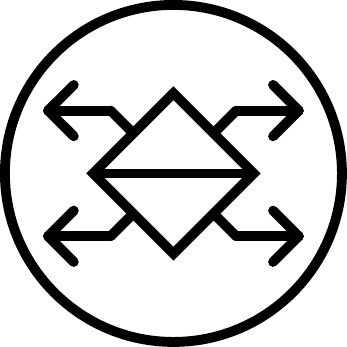}
Optical switches (nanomechanical or otherwise) can be incorporated into the above node types, but they can also stand alone in the network, switching photons from link to link without measuring them.
%\end{description}

This list is by no means exhaustive but covers the main components of a quantum network.
The division into end, repeater and support nodes is not mutually exclusive, as there may be some overlap in functionality.
For example, the ABSA may be viewed as a type of repeater node as well, as it realizes the task of entanglement swapping. The ABSA requires sophisticated RuleSets and is visible in the connect planning process; the simpler BSA, on the other hand, is tasked only with notifying two nodes about the success of entanglement creation, and need not be visible to nodes farther away in the path.
%On the other hand it is not currently equipped to handle error management in the form of purification which, of course, could change in the future.

\subsection{Routing}
\label{sec:routing}

Routing is the process of determining the path of communication between a given set of end nodes. In quantum networks, there are two distinct routes used: one that consists of quantum nodes, and a separate set of classical routes between the control mechanisms of each of those quantum devices.

%\michal{}
Picking a route can be achieved with qDijkstra (quantum Dijkstra's algorithm)~\cite{van-meter:qDijkstra}.
The link cost in this case is defined as ``seconds per Bell pair of some index fidelity $F$''. 
% We discuss in Sec~\ref{sec:reasoning_with_RS} how 
Fidelity is not an easy metric to obtain in practice, and requires constant link monitoring. An expensive but accurate measure is via \emph{tomography} of the link; lower-cost means of characterizing quantum states is an active area of research~\cite{eisert2020quantum}.
%however in the context of link cost it makes a lot of sense and therefore is worth the trouble of obtaining it.
%High data rate but poor fidelity may be less useful than low data rate but high fidelity.
%If the base fidelity is too low, purification needs to be taken into account, which automatically reduces throughput by half and more likely by 3/4 or more.
By including fidelity in the link metric, route calculation automatically takes into account the tradeoff between links with high data rate but poor fidelity versus those with low data rate and high fidelity. This approach has yielded good agreement between calculated path cost and throughput obtained via simulation of various paths with heterogeneous links~\cite{van-meter:qDijkstra}.

One of the big open questions that we are investigating is how to combine paths with multiplexing and resource reservation (and starvation), which we take up next.

\subsection{Multiplexing and Resource Reservation}
\label{sec:muxing}

%\michal{}
Circuit switching, time-division multiplexing, statistical multiplexing (like Internet best-effort forwarding) and buffer space multiplexing are all possible approaches. In buffer space multiplexing, each qubit at each router or repeater node is assigned to one of the specific connections passing through the node, akin to network slicing~\cite{barakabitze2020106984}. Aparicio studied aggregate throughput and fairness for these approaches, and found that statistical multiplexing works pretty well~\cite{aparicio2011master,aparicio11:repeater-muxing}.
Statmux allows separate regions of the network to work productively at the same time while sharing the bottleneck link, surpassing circuit switching in terms of  aggregate throughput.
However, those simulations were for small-scale networks. We believe this topic needs to be studied in much more detail to assess robustness in the face of complex, varying traffic patterns.  In particular, we fear that something akin to congestion collapse is possible, or that short-distance connections can starve long-distance connections.
%We already know that connection state will have to be maintained at repeaters and routers; quite probably there will have to be some active management of resources here, as well.

Multiplexing has to coordinate with routing and with AAA, below.  Naturally, we want to avoid a fully blocking multiplexing protocol if possible. Any multiplexing scheme that results in extended occupation of resources requires us to determine how those resources are to be allocated, and such a policy will involve identity and likely some form of payment or at minimum debit against some system credit.

\subsection{Authentication, Authorization and\\ Accounting}
\label{sec:aaa}

As just noted, it seems likely that performance well below demand will force early implementations to adopt fixed allocation of resources to individual connections. This, in turn, implies that authentication, authorization and accounting (AAA) will become important elements of the architecture \cite{rfc6733}.

Economics may come to define who has access to the early networks, unless an AAA architecture that explicitly focuses on fairness or some metric other than direct bids for access is put into place.
%Systems like the traditional operating system multi-level feedback queue or complex queueing mechanisms can also be envisioned here~\cite{cho1999managing}.

\subsection{Security}
\label{sec:security}

% \rdv{Satoh}

%\michal{}
Quantum mechanics promises unprecedented levels of confidentiality between communicating parties, which is why quantum key distribution has attracted attention of the theoretical physics and computer science community.
However, the focus on QKD also painted a skewed and incomplete picture of security in quantum networks as a whole.
This has been slowly changing lately and it has been recognized that while in principle quantum mechanics offers new methods of detecting malicious players in a network, it also enables new vectors of attack~\cite{satoh2021attacking}.

All of the protocols discussed above need authentication and tamper resistance; whether privacy is also required or useful is an open question.
Given the previous Internet (and, to a lesser extent, telephone network) experiences with lack of security in routing, accounting, etc., and the likely high cost of quantum connections, it is imperative to have a solid framework in place very early in the Quantum Internet, ideally well before a truly operational network is implemented.
This ties into the multiplexing and AAA decisions as outlined above.
%-- you can't have accounting and authorization without authentication.
%\michal{Should we stick to ``Triple-A'' or ``AAA'', or it does not matter?}

% Bringing us to point nine, network security: in fact, we are the only group in the world looking at security of network operations for quantum repeaters. But this doesn't mean at all that we have a complete plan for secure operation of the Quantum Internet.  Fairly obviously, all of the protocols we've talked about above need authentication and tamper resistance; whether privacy is also required or useful is an open question.  Given the previous Internet (and, to a lesser extent, telephone network) experiences with lack of security in routing, accounting, DoS, etc., and the likely high cost of quantum connections, it seems pretty imperative to have a solid framework in place very early in the Quantum Internet, basically well before we have a truly operational network.  And, this ties into the muxing decisions as outlined above -- you can't have accounting and authorization without authentication.

\section{Internetworking and Scalability: Recursion}
\label{sec:recursion}

%---------------------------
\begin{figure}[!t]
    \centering
    %\adjustbox{margin=1em,width=\linewidth,set height=4cm,set depth=4cm,frame,center}{QRNA}
    \includegraphics[width=0.9\columnwidth]{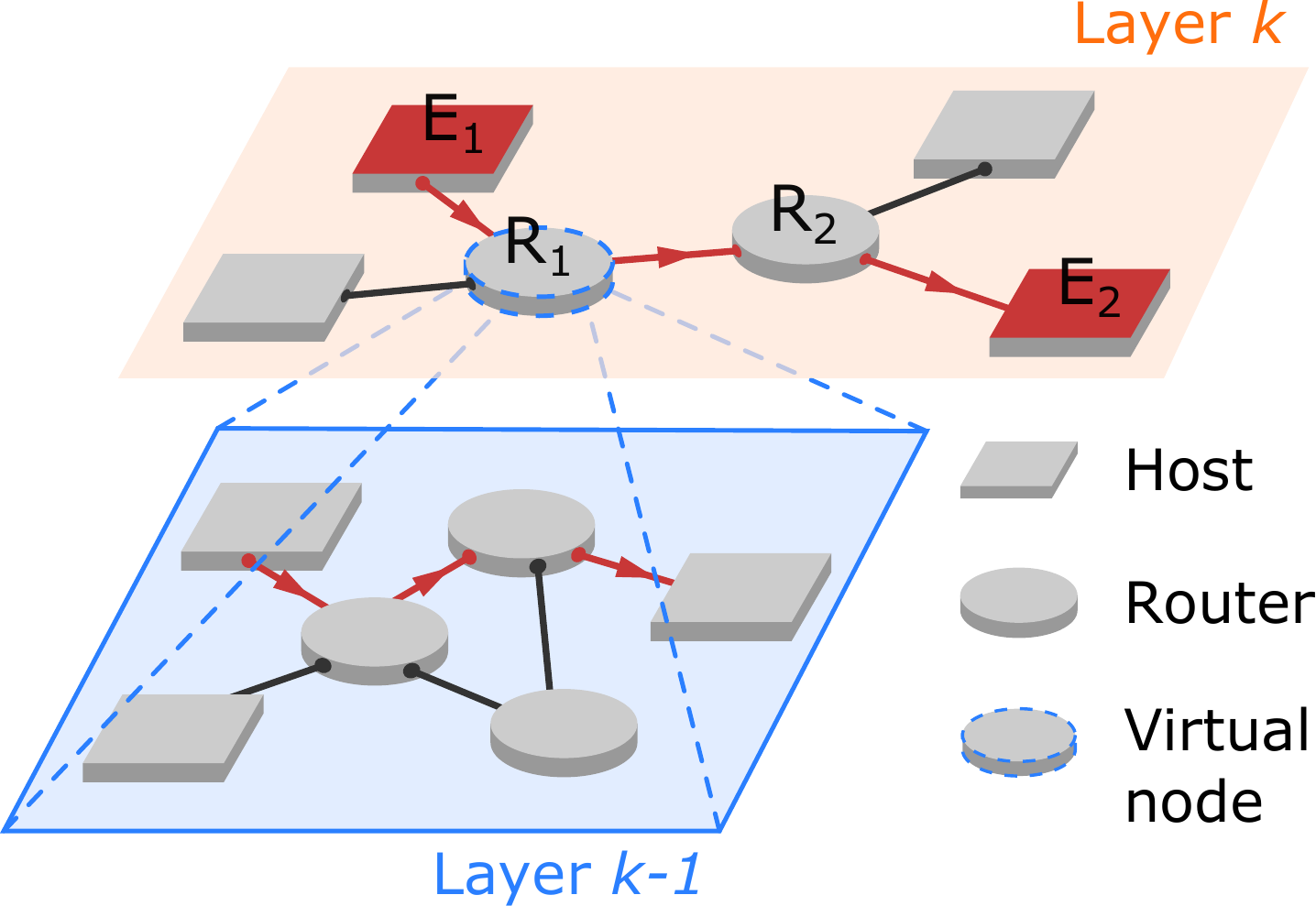}
\caption{\label{fig:qrna} QRNA uses a fully recursive architecture that can virtualize a network as a node.  Note that QRNA can work down to the link layer, or networks can be internally different~\cite{kozlowski00:qnp} as long as they participate at the network border.}
\end{figure}
%---------------------------

%\rdv{This needs to echo Secs. 3 and 4 a little better; after all, this is supposed to be an internetworking paper. Show how the RSes extend to internetworking, and how this works with the other topics in 4 (AAA, security, routing).}

An idealization of today's Internet is that it is a two-level system.
External gateway protocols such BGP are used for routing between networks while internal gateway protocols such as OSPF and IS-IS are responsible for routing within the networks.
The reality, however, is not so elegant.
Tunneling, switched Ethernets requiring spanning tree protocol underneath even though they are nominally ``link-layer'', and recent emphasis on virtualization of networks and services~\cite{barakabitze2020106984} has shaped the Internet into a multi-tier system with ad hoc interactions at each level.
Given the opportunity to create the system from scratch, and knowing the evolution path that the Internet has taken, we would probably design the Internet in a unified way that naturally takes into account interactions across multiple layers.

One such unified approach, known as the Recursive Network Architecture, was proposed by Touch \emph{et al.}~\cite{touch2006recursive}.
RNA presents an attractive blueprint for the design of the Quantum Internet, which Van Meter \emph{et al.} named the Quantum Recursive Network Architecture (QRNA)~\cite{vanmeter2011recursive}. This approach is intended to provide scalability to global proportions, including connecting physically and logically heterogeneous networks and providing autonomy, security and privacy.

Recursion naturally affects naming (Sec.~\ref{sec:naming}) and routing (Sec.~\ref{sec:routing}).
Recursion describes the hierarchy of names; the relationship among names can be described as a directed acyclic graph. This approach provides scalability in naming and routing, and enhances autonomy, security and privacy.
%\touch{it doesn't affect it per se (i.e., if the directed acyclic graph of names doesn't exist, recursion can't either. I don't know that names have anything to do with statefulness per se}
% \rdv{This isn't well explained yet.}

% If connections are \emph{stateful}, requiring (classical) state at each router, this is also affected by recursion. In the datagram model Internet, when a packet arrives at a border router, it implicitly requests that it be forwarded across the network to an exit gateway (for transit) or to the matching end node (for termination). While end nodes carry connection and application state, routers do not, at least in theory.

%\touch{IMO, both classical and quantum have the same property of distributed computation in a sense; for classical, in both cases the computation establishes shared state. what differs is whether the state has superposition and is entangled or not} \rdv{True for the end nodes, but not the mid-path nodes in the Internet, in the abstraction. Circuit switched systems or hard reservation systems do require router state.}

%\michal{}
Traditionally, connections may be of two types; boundary-to-boundary for transit and boundary-to-end node for termination.
In QRNA, both of these connections are treated as the same thing but at different levels of the network.
In Fig.~\ref{fig:qrna}, a host node $E_1$ wishes to establish a end-to-end connection with another host node $E_2$ at Layer $k$.
From the perspective of Layer $k$ the path to $E_2$ is straightforward and leads through routers $R_1$ and $R_2$.
When the connection request reaches the first router it is embedded and passed to Layer $k-1$ by the border router.
The border router is then responsible for requesting an end-to-end connection across Layer $k-1$ to an appropriate border router that then passes the original connection request up to Layer $k$.
The recursive nature of the architecture allows the connection requests to be embedded into as many levels as is required.

\subsection{Connection Setup: Two-Pass with Rewrite}

\begin{figure}
\begin{center}
\includegraphics[width=\columnwidth]{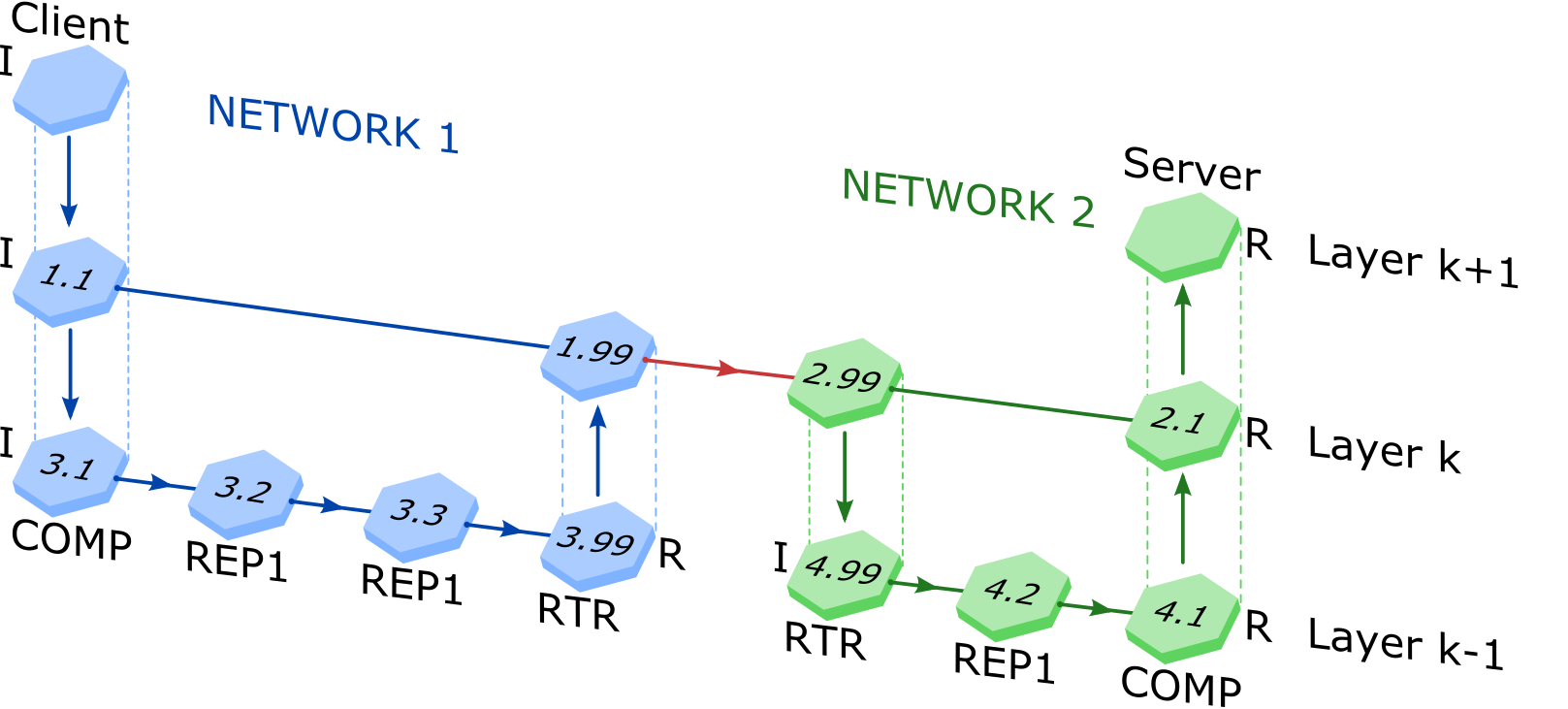}
\end{center}
\caption{\label{fig:recursive-setup} Two-pass connection setup in an internetwork. The arrows indicate how the initial connection request gets passed up/down the Layers and between the networks. Each Layer has its own Initiator (I) and Responder (R), due to the recursive encapsulation of the connection request Layer $k-1$ has two pairs I-R.
% At Layer $k$, the path is three hops: E$_1\rightarrow$ RTR1$\rightarrow$ RTR2$\rightarrow$E$_2$. At Layer $k-1$, the total path consists of more hops, but RuleSets are constructed only end node to border or border to border.  The border gateway router 1.99, which acts as Responder to the network to the left, is a distributed innovation point, as is the end-node Responder (2.1) on the right. The recursive rewriting process between layers, highlighted in red, is an open topic of research.
}
\end{figure}

Recursion must work with the two-pass connection setup described in Sec.~\ref{sec:two-pass}. We accomplish this via RuleSet rewriting where crossing recursion layers, such as at network boundaries.
Setup in an internetwork is shown in Fig.~\ref{fig:recursive-setup} (compare to Fig.~\ref{fig:simple-setup}).  In order to maintain network autonomy and privacy and improve scalability, the border router rewrites the existing set of link information into a single hop, much like a single hop in BGP routing hides network internal topological information for the same purposes. The border router acts as a Responder to the original Initiator, and its estimate of the performance of the path from the Initiator to its location is used to derive the performance characteristics it reports when describing the virtual link at a higher layer of recursion.

%---------------------------
%% %---------------------------

In the Quantum Internet (at least through the first two generations), a connection is a form of distributed computation, with all nodes on the path participating in purification and entanglement swapping.  Connections will have to be established in advance along the path, and will be \emph{stateful}.
% In Fig.~\ref{fig:recursive-setup}, 1.2 and 1.3 hold RuleSets only for Level $k-1$, as created and distributed by the border router 1.99. Level $k$ RuleSets are E2E, built on the virtualized links presented by Level $k-1$.

During connection setup, at every layer, the node is given a Responder (destination) address and can determine the \verb|nexthop| based on local selection policy. To advance the setup process, the node sends the request to the neighbor (if we have reached the physical link level) or recurses. At layer $k$, we recurse to layer $k-1$ by translating our $k$ address and the $k$ layer \verb|nexthop| to layer $k-1$ addresses, then passing to layer $k-1$ with the latter as the new Responder address.
Each network constructs provisional RuleSets upon the connection request reaching the corresponding Layer $k-1$ Responders (3.99 and 4.1 in Fig.~\ref{fig:recursive-setup}). These RuleSets are distributed backwards along the network path (not shown in Fig.~\ref{fig:recursive-setup}). Upon acceptance of the connection request by the Server, a reply is sent backwards along the same path confirming the RuleSets.

Performing recursion at administrative boundaries has several benefits: a) it limits the amount of information each node has to have on hand about the entire internetwork, enhancing E2E scalability and network autonomy; b) it allows Responders to innovate (within the bounds of the RuleSet architecture); c) it allows us to reason about connections effectively; d) it serves as a convenient point for 1G-2G inter-operation as new technologies are deployed~\cite{PhysRevA.93.042338}; and e) it facilitates interoperation with different network architectures~\cite{kozlowski00:qnp}.

% \naphann{Do we really need to separate the 1G and 2G into a different network altogether? Isn't the idea of 2G is just adding error corrected codes over a physical link? The link still need heralded entanglement generation (or maybe EPPS?), so 2G can be done using RuleSet over the same technology as 1G but just need more qubits for the encoding, (and/or higher base fidelity or we just purify it enough times at the link level?). But other than that aren't they the same? }
%\rdv{That’s a very interesting question that has only recently crossed my mind as we have been working on detailed RuleSets for the last six months or so. Prior to that, I had always assumed that a give network would be either 1G or 2G. I still think that’s more likely; the demands that 2G networks make on hardware and software are so much more complex that I don’t think it will be easy to deploy incrementally within a network. I think it’s a lot more likely to be deployed as a new network, but one that is required to interoperate at a boundary with the old network. But it’s a little hard to see for certain...}

\subsection{Routing, Multiplexing and AAA}

The core routing problem of selecting a path in Internet-scale systems is solved, as noted above, using two-level or three-level systems, with the Internet's top level being the global BGP. However, issues of policy, economics and especially of security still exist at the top level~\cite{rfc7454}. In QRNA, this approach is generalized and extended using full recursion; a routing protocol is required at each layer.

At the lowest layer, we follow the qDijkstra link cost metric of seconds per Bell pair at a particular fidelity. Using QRNA's recursion, at the next layer up, the intra-network path will appear as a link. This link will, in turn, have a reportable performance metric of Bell pair creation rate. However, as the intra-network RuleSet can be tuned with different numbers of rounds of purification, that rate can be traded off for higher fidelity. Prior work has shown that performing more purification closer to the link level results in higher end-to-end throughput~\cite{van-meter07:banded-repeater-ton}, so we expect the policy to be set such that each network presents a slower but higher quality link.

A bigger problem is multiplexing, which as noted requires AAA. Any network will have many connections originating, terminating or transiting. The Internet community unfortunately provides less guidance here; inter-domain QoS mechanisms have been under development since the 1990s but are not widely deployed. Thus, we consider this to be one of the most important open research issues.

%\input{implementation}
%---------------------------

\section{Evidence}
\label{sec:evidence}
%---------------------------
\if0
\begin{figure}[!t]
    \centering
    \includegraphics[]{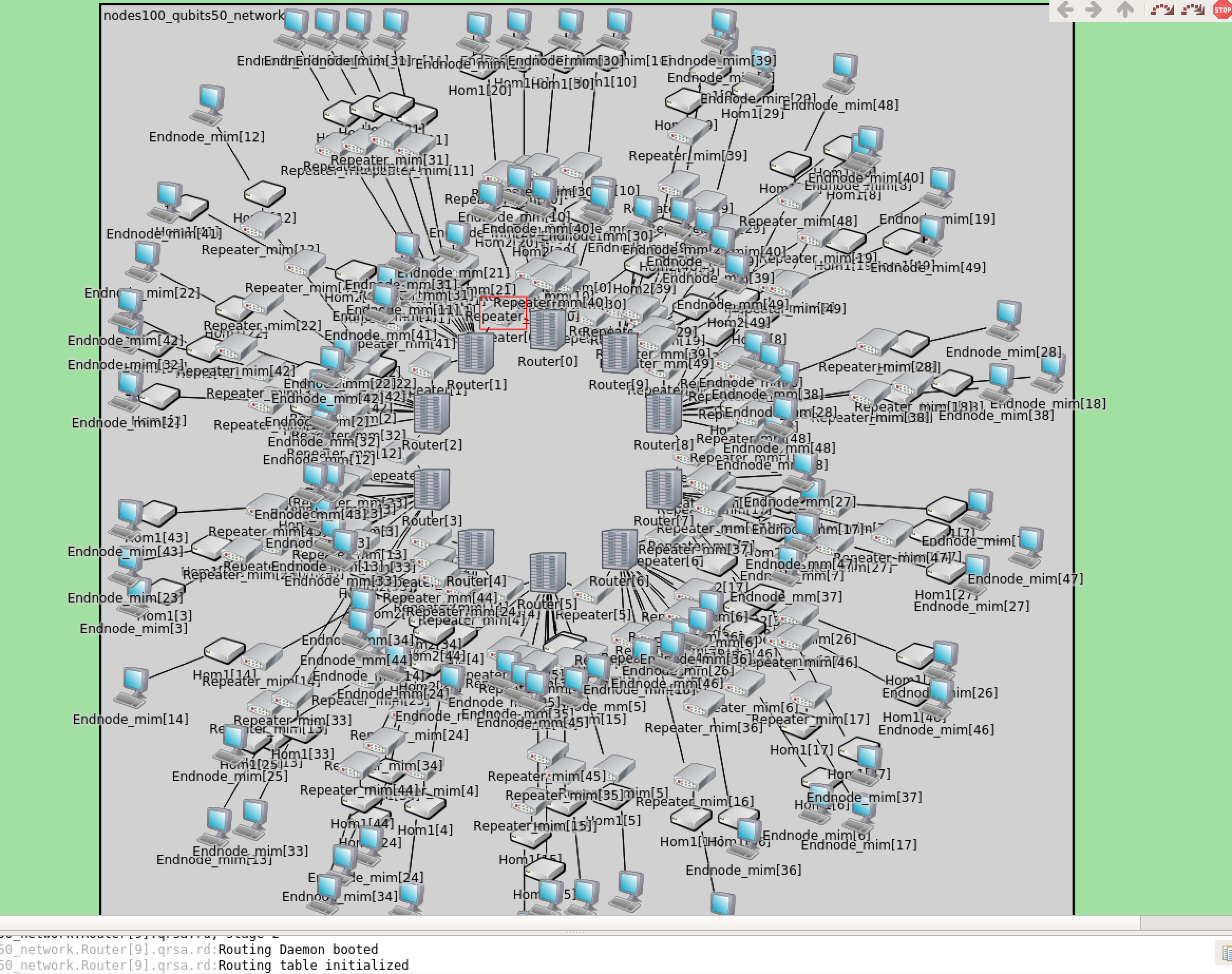}
    %\adjustbox{margin=1em,width=\linewidth,set height=4cm,set depth=4cm,frame,center}{Evidence that QRNA is glorious.}
\caption{\label{fig:proof1} \rdv{Evidence figure.}}
\end{figure}
\fi
%---------------------------
\begin{figure}[!t]
    \centering
    \includegraphics[width=0.9\columnwidth]{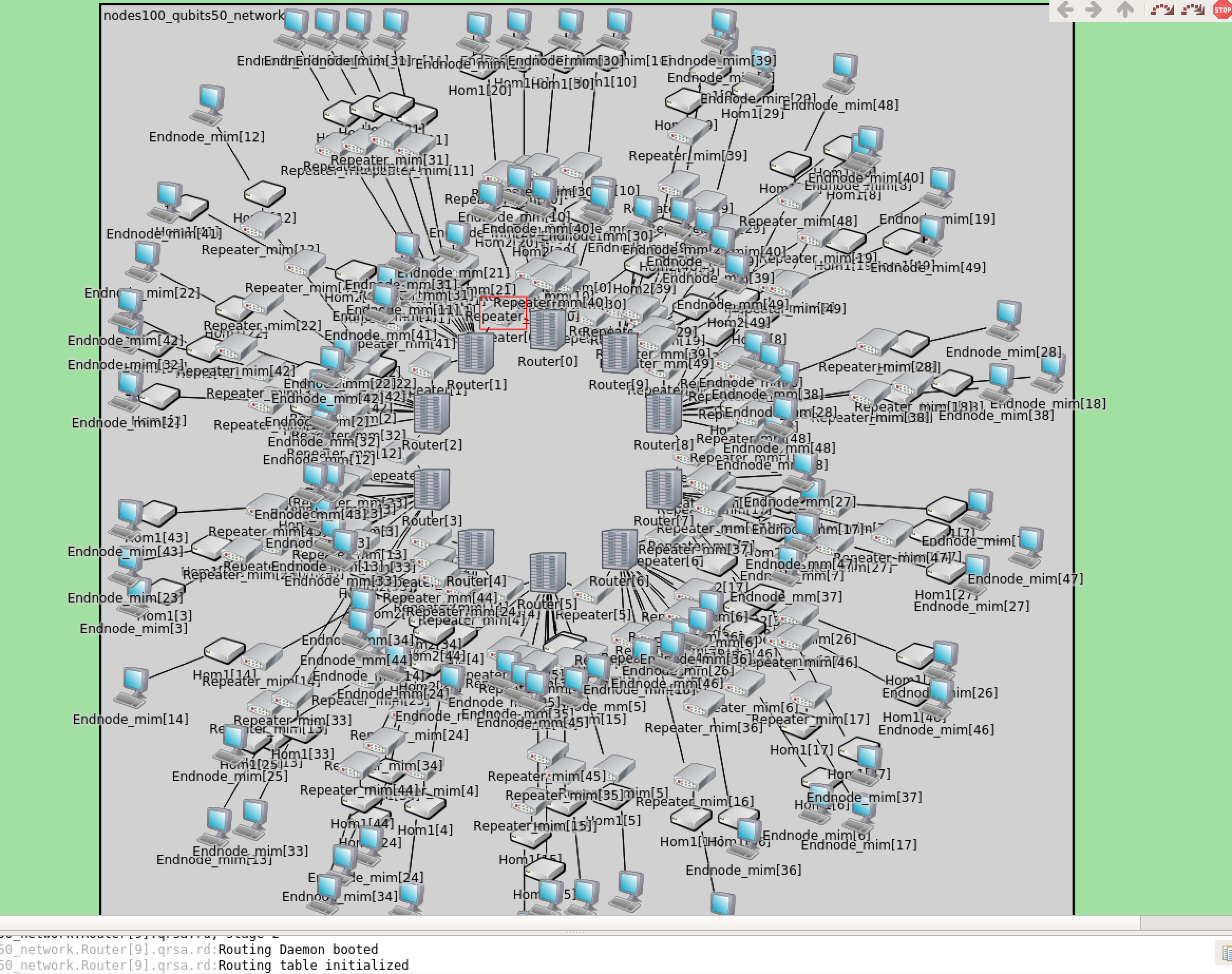}
    %\adjustbox{margin=1em,width=\linewidth,set height=4cm,set depth=4cm,frame,center}{QuISP is a form of evidence.}
\caption{\label{fig:quisp} Our open source simulator, QuISP, focuses on protocol and scaling issues in order to further network and internetwork architecture research. Here, 100 COMP, 110 REP1, and 10 RTR nodes, collectively having 44,000 memory qubits, are connected via 220 links. 100 of the links use BSA nodes; the rest are direct connections.}
\end{figure}
%---------------------------

%\rdv{I think I'd like to redo the basic figure showing the relationship between qDijkstra cost and E2E throughput. That means we need full ES and purification. (Do we do multiple purification schemes?) Can be tested on a moderate-size, heterogeneous network, as long as we have enough heterogeneity in links.  Means we need a decent scheme for picking ES points for various paths of varying distances and hop counts.}

In an ideal world, the long-term proof of an architecture would be widespread adoption. In reality, of course, plenty of splendid architectures (processors, OSes, communication systems) have fallen by the wayside for reasons unrelated to technology.  Moreover, such a retrospective view does not help us assess a prospective architecture. Here, we discuss how the RuleSet approach leads to robust protocols, and how we are validating our architecture via simulation, documenting the protocols, and working toward real-world implementation.

\subsection{Reasoning with RuleSets}
\label{sec:reasoning_with_RS}
%---------------------------
\if0
\begin{figure}[!t]
    \centering
    \adjustbox{margin=1em,width=\linewidth,set height=4cm,set depth=4cm,frame,center}{State machines and RuleSets}
\caption{\label{fig:reasoning} \rdv{Show how state machines and RSes correspond?}}
\end{figure}
\fi
%---------------------------

% \rdv{Show how state machines and RSes correspond? Or will folks w/ SDN knowledge consider that obvious?}

% \naphann{As I mentioned it at Condition Clause section, I don't think RSes and state machines correspond. We could in some way say that the progression of resources follow the state machine diagram but the RS as a whole isn't since every Rule runs in parallel}

% \rdv{Should either or both of these have figures?}

A key purpose of the RuleSet architecture is to make it possible to reason rigorously about distributed behavior. At any point in time, we can enumerate the set of possible events at all nodes and ask if execution of specific Action Clauses will result in unwanted operation, such as \emph{leapfrogging}.  In leapfrogging, in a chain of nodes A-B-C-D, if B and C are each tasked with performing entanglement swapping, uncoordinated selection of resource states can result in A-C and B-D entanglement, rather than the desired A-D entanglement.

In another example, if A-B entanglement has been achieved and B is waiting on B-C entanglement to perform swapping, a \emph{race condition} can occur in which A decides to discard the Bell pair (due to memory decoherence incurred during a long wait) just as B receives notification of B-C entanglement and performs the swapping operation. The message from B to A informing A of the swapping event arrives too late, and A has already reinitialized its qubit for reuse.  This is especially problematic if C chooses, upon receipt of notification from B, to use the ersatz A-C pair to teleport C's important data to A.  Using RuleSet logic, we can detect this potential race condition and define Rules such that A will not discard the Bell pair until after B does, by giving B a discard timer that is more than the one-way messaging latency with A.
%\rdv{This is an easy diagram if we have room.}

%\rdv{This paragraph might belong somewhere else.}
% A considerable problem in reasoning with RuleSets is the desire to use state fidelity in Condition Clauses. While this seems to be an obvious approach to efficient end-to-end operation, we have the challenge that \emph{measuring fidelity is expensive, but predicting fidelity is difficult}~\cite{eisert2020quantum,nishio20:error-aware}. 

% RuleSets that make use of Fidelity Condition Clauses may be feasible only with high-quality fidelity estimation based on known characteristics of hardware paths, which is theoretically straightforward but difficult in practice, especially when entangled states involving large numbers of qubits are the subject. Thus, in practice, a better approach may be RuleSets that are more rigid, focusing on conservatively-chosen timers and fixed sequences of purification and swapping operations.

% \subsection{Towards Implementation}
% \subsection{Simulation}
\subsection{Simulation, Specification and\\ Implementation}

To validate our designs, we are implementing a highly scalable simulator called QuISP (Quantum Internet Simulation Package)~\cite{satoh21:_quisp}. 1G networks, entanglement swapping and purification governed by RuleSets, and connection setup are complete (but continue to evolve); rudimentary routing and circuit switched multiplexing are all functional and pass included tests, but remain in active development. All-optical paths are in active development. RuleSets are currently being designed for 2G and multi-party states. The full QRNA protocol set is in design, and the simulator's performance has been measured to scale adequately for hundreds of nodes on a laptop, enough to demonstrate complex, multi-level, recursive internetworking.

% \subsection{Specification}

Any network system, especially one intended to be open, must be supported by specifications for protocols and behavior. The difficulty of writing such documents can be viewed as one piece of evidence about the elegance and simplicity of an architecture. Our simulator work began with a set of design documents, and we have specifications for some of the core protocols now in development. %\rdv{We need to make this be more true about specs.}

Moreover, we are working closely with the Quantum Internet Task Force (QITF), a quantum Internet testbed initiative that expects to build not only a single network but to actually focus on scalability in network and internetwork architecture. We expect some aspects of the architecture presented here to be adopted directly, while others doubtless will undergo significant evolution as a result of the collaboration.
\section{Conclusion}
\label{sec:conclusion}

Ultimately, our proposed quantum internetwork architecture builds on three critical points: a recursive architecture for internetworking and scalability, RuleSet-based connection operation providing the right vocabulary across disparate hardware, and a two-pass connection setup routine (outbound info collection, inbound RuleSet distribution). This structure will allow for continuing evolution of the internetwork, providing a platform for distributed, independent advances in physical technology and in protocols.

Our work is maturing rapidly, with design, specification, and simulation well advanced and real-world implementation in the serious planning stages. In particular, with different groups now involved in detailed discussions, the RuleSet design will be challenged to work in heterogeneous environments, which we expect to further validate the general approach even as it is likely that details will change. Although there is solid work on routing and multiplexing, designing a system that will be robust at scale and that will serve us well for decades is perhaps the area of most concern. 

With this structure in place, we feel that architecture and protocols are on pace to mature to usable levels alongside hardware, though as noted in the introduction experience shows that architecture matures more slowly. However, we expect to take full advantage of knowledge gained over the last half-century of data networking research and development. This should carry us through evolutionary stages to a full Quantum Internet supporting cryptographic, sensor, and distributed computation applications.
%\input{acks}

%-------------------------------------------------------------------------------
\section*{Acknowledgments and Availability}
%-------------------------------------------------------------------------------

This material is based upon work supported by the Air Force Office of Scientific Research under award number FA2386-19-1-4038.

The authors thank Joe Touch for clarification of past contributions.

Our open source simulator~\footnote{\url{https://github.com/sfc-aqua/quisp##quisp}} and in-preparation RFC-like specifications are or will be made available on the web.

%-------------------------------------------------------------------------------
\clearpage
%\rdv{DOIs and URLs don't show up cleanly here.}
\bibliographystyle{plain}
\bibliography{arch.bib}

\clearpage
\appendix
\section{Quantum Concepts}
\label{sec:qconcepts}
% \rdv{cribbed from WIDE reports}

There are many good introductions to quantum computing, on the web~\cite{futurelearn} and in print~\cite{sutor19:dancing}, but for convenience the following is a brief summary of the key aspects of quantum
communication and computation that impact network and system
architecture.

The primary difference between quantum mechanics and classical probability is that quantum mechanics uses \emph{probability amplitudes}, rather than straight probabilities~\cite{aaronson2013quantum}.  Probability amplitudes can be complex numbers; if the amplitude of a given state is $\alpha$, then the probability of finding that state is $|\alpha|^2$. Most of the concepts below derive fairly directly from this fact and the general wave nature of quantum systems.

Quantum information is most often discussed in terms of
\emph{qubits}. A qubit, like a classical bit, is something with two possible values that we can label zero and one.  Unlike a classical bit, a qubit can occupy both values simultaneously, known as \emph{superposition}.

To understand quantum computation, we need seven basic concepts:

{\bf Superposition.}  A qubit can represent multiple
values in different proportions at the same time, e.g., two-thirds of a ``one'' and one-third of a ``zero''.  This
\emph{superposition} determines the relative probability of finding each value when we \emph{measure} the state.

{\bf Entanglement (and Bell pairs).}  Groups of qubits can exhibit strong correlation between the qubits that cannot be explained by independent probabilities for individual qubits.  Instead, the group must be considered as a whole, with interdependent probabilities.  This phenomenon is known as \emph{quantum entanglement}.  A special entangled state known as a {\em Bell pair} or \emph{EPR pair}, consisting of two quantum bits, figures
prominently in quantum communication.  Each qubit in the pair has a 50\% probability of having a value of 1 and a 50\% probability of having a value of 0 when we measure it.  Although we cannot predict which will be found, when we measure one member of the pair, the value
of the other is immediately determined.  This happens independent of the distance between the two members of the Bell pair.

{\bf Interference.} Quantum algorithms use some building blocks derived from classical concepts, such as adder designs, but the overall thrust of a quantum algorithm is very different from that of a classical algorithm.  Rather than attempting to solve a problem and checking for the answer, a quantum algorithm's goal is to create
\emph{interference} between the elements of a superposition quantum state.  Constructive interference reinforces desirable states, increasing the probability of finding a desirable outcome on measurement, while destructive interference reduces the probability.

{\bf Unitary, or reversible, gates.} Manipulating those probability amplitudes, including creating entanglement and making the interference patterns, involves the use of logical operations known as \emph{gates}. These gates are similar to Boolean logic, but must be reversible, which in mathematical terms means they are represented by a \emph{unitary} transformation matrix.

{\bf Measurement.} As described above, when we measure a qubit, we get only a single classical bit of information (the ``one'' or ``zero''), and the superposition \emph{collapses}.  The probability of finding a zero or a one depends on the probability amplitudes.

{\bf Decoherence.} Unfortunately, any physical operation (including simply storing a qubit) gradually degrades the state. Decoherence is the single most important technological fact driving quantum computer and quantum network implementations.  We can counter this by using a form of error correction or detection.

{\bf No cloning.}  As mentioned above, a key restriction of quantum systems is that we cannot make \emph{independent} copies of an unknown
state~\cite{wootters:no-cloning}.  This makes error correction difficult.

A few additional concepts will augment understanding quantum networks.

{\bf Fidelity.}  The quality of a quantum state is described by its \emph{fidelity}, which is, roughly, the probability that we correctly understand the state -- if we ran the same experiment many times and measured the results, how close to our desired statistics would we be? This is one simple measure of the amount of decoherence.

{\bf Purification.}  The form of error detection historically favored
in quantum repeater networks is {\em purification}, which uses minimal
resources~\cite{briegel98:_quant_repeater}.  It sacrifices some
quantum states to test the fidelity of others.  There are various
purification mechanisms, with different purification algorithms and
different methods for determining which states are sacrificed, each
with particular tradeoffs.

{\bf Quantum error correction (QEC).}  QEC may be based on classical
codes or purely quantum concepts.  The primary difficulties are
extraction of errors without damaging quantum state, avoiding error
propagation, and the increased resources required.  (See references
contained in \cite{van-meter12:_q_internetworking},
\cite{PhysRevA.79.032325} and \cite{PhysRevLett.104.180503}.)

{\bf Teleportation.}  Teleportation destroys the state of a qubit at the sender and recreates that state at the destination, teleporting information rather than matter~\cite{bennett:teleportation}.  The process
uses a Bell pair's long-distance correlation, followed by transmission of a pair of classical bits.  Teleportation consumes a Bell pair.

{\bf Entanglement swapping.}  Splicing two long-distance Bell pairs together to make one longer Bell pair is known as entanglement swapping.

With these basic concepts, we can begin to construct networks.
%Bell pairs are consumed by teleportation, so one way to organize a network is to create a continuous stream of Bell pairs between source and destination -- as long as we identify those sources and destinations, choose paths to get there, and manage the resources along the way.
For those interested in a more research-oriented, in-depth survey of quantum computing systems, we recommend the following short list of papers:~\cite{divincenzo2000piq,chong2017programming,devitt13:rpp-qec,harrow2017quantum,ladd10:_quantum_computers,montanaro2015:qualgo-qi,preskill2018quantum,van-meter13:_blueprint,van-meter19:_quant_tweet_zen}. For communication, we recommend:~\cite{Wehner18:eaam9288,I-D.irtf-qirg-principles,I-D.irtf-qirg-quantum-internet-use-cases,PRXQuantum.2.017002,briegel98:_quant_repeater,kimble08:_quant_internet}.

%\input{intangibles_table}

%%%%%%%%%%%%%%%%%%%%%%%%%%%%%%%%%%%%%%%%%%%%%%%%%%%%%%%%%%%%%%%%%%%%%%%%%%%%%%%%
\end{document}